\documentclass[a4paper,UKenglish]{lipics-v2018}
\nolinenumbers
\usepackage{microtype}%if unwanted, comment out or use option "draft"
\usepackage{tcolorbox}
\usepackage{amssymb}
\usepackage{amsbsy}
\usepackage{amsmath}
\usepackage{bbm}
\usepackage{latexsym}
\usepackage{xspace}

\title{Effective Divergence Analysis for Linear Recurrence Sequences}%\footnote{Jo\"el Ouaknine was supported by ERC grant AVS-ISS (648701), and James Worrell was supported by EPSRC Fellowship EP/N008197/1.}} 

\author{Shaull Almagor}{Department of Computer Science, Oxford University, UK}{shaull.almagor@cs.ox.ac.uk}{}{}
\author{Brynmor Chapman}{MIT CSAIL}{brynmor@mit.edu}{}{}
\author{Mehran Hosseini}{Department of Computer Science, Oxford University, UK}{mehran.hosseini@cs.ox.ac.uk}{}{}
\author{Jo\"el Ouaknine}{Max Planck Institute for Software Systems,  
	Germany \& \\Department of Computer Science, Oxford University, UK}{joel@mpi-sws.org}{}{Supported by ERC grant AVS-ISS (648701)}
\author{James Worrell}{Department of Computer Science, Oxford University, UK}{jbw@cs.ox.ac.uk}{}{Supported by EPSRC Fellowship EP/N008197/1}

\authorrunning{S.\,Almagor, B.\,Chapman, M.\,Hosseini, J.\,Ouaknine and J.\,Worrell}%mandatory. First: Use abbreviated first/middle names. Second (only in severe cases): Use first author plus 'et al.'

\Copyright{Shaull Almagor, Brynmor Chapman, Mehran Hosseini, Jo\"el Ouaknine, and James Worrell} %mandatory, please use full first names. LIPIcs license is "CC-BY";  http://creativecommons.org/licenses/by/3.0/

\subjclass{\ccsdesc[500]{Computing methodologies~Algebraic algorithms}, \ccsdesc[500]{Theory of computation~Logic and verification}}% mandatory: Please choose ACM 2012 classifications from https://www.acm.org/publications/class-2012 or https://dl.acm.org/ccs/ccs_flat.cfm . E.g., cite as "General and reference \(\rightarrow\) General literature" or \ccsdesc[100]{General and reference~General literature}. 

\keywords{Linear Recurrence Sequences, Divergence, Algebraic Numbers, Positivity}%mandatory

\category{}%optional, e.g. invited paper

\relatedversion{}%optional, e.g. full version hosted on arXiv, HAL, or other respository/website

\supplement{}%optional, e.g. related research data, source code, ... hosted on a repository like zenodo, figshare, GitHub, ...

\funding{}%optional, to capture a funding statement, which applies to all authors. Please enter author specific funding statements as fifth argument of the \author macro.

\acknowledgements{}%optional

\EventEditors{Sven Schewe and Lijun Zhang}
\EventNoEds{2}
\EventLongTitle{29th International Conference on Concurrency Theory (CONCUR 2018)}
\EventShortTitle{CONCUR 2018}
\EventAcronym{CONCUR}
\EventYear{2018}
\EventDate{September 4--7, 2018}
\EventLocation{Beijing, China}
\EventLogo{}
\SeriesVolume{118}
\ArticleNo{42}
\theoremstyle{plain}
\newtheorem{proposition}[theorem]{Proposition}
\newtheorem{property}[theorem]{Property}

\newcommand{\tup}[1]{\langle #1 \rangle}

\renewcommand{\phi}{\varphi}

\newcommand{\set}[1]{\left\{#1\right\}}
\newcommand{\NN}{{\mathbb{N}}}
\newcommand{\ZZ}{{\mathbb{Z}}}
\newcommand{\RR}{{\mathbb{R}}}
\newcommand{\CC}{{\mathbb{C}}}
\newcommand{\QQ}{{\mathbb{Q}}}
\newcommand{\TT}{{\mathbbm{T}}}

\newcommand{\re}{{\rm Re}}
\newcommand{\im}{{\rm Im}}
\newcommand{\conj}[1]{{\overline{#1}}}

\newcommand{\norm}[1]{{\left\lVert#1\right\rVert}}

\newcommand*\Bell{\ensuremath{\boldsymbol\ell}}

\newcommand{\Ptime}{{\bf PTIME }}
\newcommand{\posslp}{{\bf PosSLP}}
\newcommand{\NPposslp}{\(\text{{\bf NP}}^\text{{\bf PosSLP}}\)}
\newcommand{\coNPposslp}{\(\text{{\bf coNP}}^\text{{\bf PosSLP}}\)}

\newcommand{\homLRS}{\textsc{hom}}

\begin{document}

\maketitle

\begin{abstract}
  We study the growth behaviour of rational linear recurrence
  sequences. We show that for low-order sequences, divergence is
  decidable in polynomial time. We also exhibit a polynomial-time
  algorithm which takes as input a divergent rational linear
  recurrence sequence and computes effective fine-grained lower bounds
  on the growth rate of the sequence.
\end{abstract}

\section{Introduction}
\label{sec:intro}
Linear recurrence sequences (LRS), such as the Fibonacci numbers,
permeate a wide range of scientific fields, from economics and
theoretical biology to computer science and mathematics. In
computer-aided verification, for example, LRS techniques play a key
r\^{o}le in the termination analysis of a large class of simple while
loops---see~\cite{OW15} for a short survey on this topic. Likewise,
the ergodic behaviour of Markov chains in probability
theory~\cite{AAOW15}, or the stability of supply-and-demand price
equilibria in laggy markets in economics (the so-called `cobweb
model')~\cite{Bau70} can be analysed through an examination of the
asymptotic behaviour of certain types of LRS; in particular,
instability of price equilibria corresponds precisely to
\emph{divergence} of the associated LRS\@.

In this paper, we undertake a systematic and fine-grained analysis of
the growth behaviour of rational linear recurrence sequences from the
point of view of effectiveness and complexity. In order to describe
our main results, we first require some preliminary definitions. A
sequence of real numbers \(\mathbf{u} = \tup{u_n}_{n=1}^\infty\) is said
to satisfy a \emph{linear recurrence of order \(k\)} if there are real
numbers \(a_1,\ldots,a_{k+1}\) such that
\begin{gather} 
  u_{n+k} = a_1 u_{n+k-1} + \cdots + a_{k-1} u_{n+1} + a_k u_n +
  a_{k+1} \label{eq:RECUR}
\end{gather}
for all \(n\in \mathbb{N}\).  Such a recurrence is said to be
\emph{homogeneous} if \(a_{k+1}=0\) and \emph{inhomogeneous} if
\(a_{k+1}\neq 0\).
% (In the absence of a qualifier, a given sequence can be either
% homogeneous or inhomogeneous.)
The \emph{characteristic polynomial} of the recurrence is
\begin{gather*} 
p(x) := x^k - a_1x^{k-1} - \cdots - a_{k-1}x - a_k \, .
\end{gather*}
The zeros of \(p\) are called the \emph{characteristic roots}.  A
characteristic root of maximum modulus is said to be \emph{dominant}
and its modulus is the \emph{dominant modulus}. The
\emph{multiplicity} of a characteristic root \(\gamma\) is the maximal
\(m\in \NN\) such that \((x-\gamma)^m\) divides \(p(x)\).

An LRS is said to be \emph{rational} if it consists of rational
numbers, \emph{integral} if it consists of integers, and
\emph{algebraic} if it consists of algebraic numbers. An LRS is
\emph{simple} if all of its characteristic roots have multiplicity 1,
and is \emph{non-degenerate} if no ratio of two distinct
characteristic roots is a root of unity.\footnote{For most practical
  purposes---and certainly for all of the computational tasks
  considered in this paper---LRS can be assumed to be non-degenerate,
  since any degenerate LRS can be effectively decomposed into a finite
  number of non-degenerate LRS; moreover this reduction can be carried
  out in polynomial time for rational LRS of bounded
  order~\cite{EPIW03, OW14b}.}

We say that an LRS \(\mathbf{u}\) \emph{diverges to \(\infty\)} if
\(\lim_{n\rightarrow\infty}u_n = \infty\) (technically speaking: for all
\(T \in \mathbb{N}\), there exists \(N \in \mathbb{N}\) such that, for all
\(n \geq N\), we have \(u_n \geq T\)). We also say that \(\mathbf{u}\) is
\emph{absolutely divergent} (or \emph{diverges in absolute value}) if
\(\lim_{n\rightarrow\infty}|u_n| = \infty\).

The LRS \(\mathbf{u}\) is said to be \emph{positive} if \(u_n \geq 0\) for
all \(n \geq 1\), and \emph{ultimately positive} if there is some
\(N \in \mathbb{N}\) such that \(u_n \geq 0\) for all \(n \geq N\).

A celebrated result from the 1930s, the Skolem-Mahler-Lech theorem
(see~\cite{EPIW03}), implies that all non-degenerate integral LRS are
absolutely divergent. This statement is however \emph{non-effective}
in a very basic sense: given a finite representation of a
non-degenerate integral LRS \(\mathbf{u}\), there is no known algorithm
to compute a bound \(N\) such that \(u_n \neq 0\) for \(n \geq N\). It is
also worth pointing out that the divergence assertion fails in general
for non-integral LRS\@.

The question of the so-called \emph{rate of absolute divergence} for
non-degenerate integral LRS was subsequently extensively studied;
see~\cite[Sec.~2.4]{EPIW03} for an account of some of the key results
accumulated over the last several decades. To begin with, a fairly
straightforward fact is the following: if \(\mathbf{u}\) is an algebraic
LRS of order \(k\) with dominant modulus \(\rho\), then there is an
effectively computable constant \(a\) such that, for all \(n \geq 1\),
\(|u_n| \leq a\rho^nn^k\). In the 1970s, a conjecture was formulated to
the effect that any non-degenerate integral LRS has, essentially, the
maximal possible growth rate (see the next theorem for a precise
statement). The conjecture was finally settled positively
independently by Evertse~\cite{Eve84} and by van der Poorten and
Schlickewei~\cite{PS91}:
\begin{theorem}
\label{thm: S units}
  For any non-degenerate algebraic LRS
  \(\mathbf{u}\) of dominant modulus \(\rho > 1\), and any
  \(\varepsilon > 0\), there exists a constant \(N\) such that, for all
  \(n \geq N\), we have \(|u_n| \geq \rho^{(1-\varepsilon)n}\). 
\end{theorem}

This is a highly non-trivial result making use of deep
number-theoretic tools concerning bounds on the sum of
\(S\)-units. Unfortunately, the proof is not effective, in the sense
that given \(\varepsilon > 0\), it does not provide estimates for the
corresponding value of \(N\). This effectiveness issue is described as
``an important open problem'' in~\cite{EPIW03}, where it is
furthermore suggested that any progress on the matter would likely
hinge upon substantial improvements of deep number-theoretic results,
such as Roth's theorem, the prospects of which currently appear to be
remote.

Nevertheless---and in particular for algorithmic applications in
computer science---\linebreak effectiveness is of central
importance. The sharpest known results in this vein are due to
Mignotte~\cite{Mig75} as well as Shorey and Stewart~\cite{SS83},
capping a long line of work in this area:
\begin{theorem}
\label{absolute-effective-divergence}
  For any homogeneous non-degenerate integral LRS
  \(\mathbf{u}\) of order at most \(3\) with dominant modulus
  \(\rho\), there are effective constants \(a, d\) and \(N\) such that, for all
  \(n \geq N\), we have \(|u_n| \geq \frac{a\rho^n}{n^d}\).
\end{theorem}

For most problems in computer science and automated verification, such
as the analysis of the long-run behaviour of dynamical systems or the
termination of linear while loops, the primary notion of
\emph{divergence} is clearly much more relevant than that of
`divergence in absolute value'. In view of the above results, however,
one might expect that little could be said about effective rates of
divergence. Somewhat surprisingly, divergence does turn out to be
significantly more tractable than absolute divergence. At a high
level, the main results of this paper can now be summarised as
follows:
\begin{tcolorbox}
\begin{quote}
  Given a rational LRS \(\mathbf{u}\), homogeneous or inhomogeneous,
  either of order at most \(5\), or, if the LRS is simple, of order at
  most \(8\), we can carry out the following tasks in polynomial time:
\begin{itemize}
\item decide if \(\mathbf{u}\) diverges to \(\infty\) or not; and
\item in divergent instances, provide effective fine-grained lower
  bounds on the rate of divergence of \(\mathbf{u}\).
\end{itemize}
\end{quote}
\end{tcolorbox}

The precise statements can be found in Theorems~\ref{div-1} and
\ref{div-2}. The most obvious contrast in comparison with
Theorem~\ref{absolute-effective-divergence} is the higher order of LRS
that can be handled effectively (5 and 8 versus 3). It is also worth
noting, however, that our results apply more generally to rational (as
opposed to integral) LRS, and that we can handle inhomogeneous
sequences at no cost---this is remarkable in that the folklore wisdom
usually broadly equates inhomogeneous LRS of order \(k\) with
homogeneous LRS of order \(k+1\) (this assertion, as well as the manner
in which we circumvent it, are made precise in the main body of the
paper).

Finally, let us point out that our analysis of divergence rates
relies, among others, on improvements to results concerning the
positivity and ultimate positivity of LRS, which were originally
developed in~\cite{OW14a,OW14b,OW14c}. As a by-product, therefore,
stronger results on the Positivity and Ultimate Positivity
Problems---notably dealing with inhomogeneous LRS---can be found in
the present paper, in particular in the form of Theorems~\ref{thm:
  ultimate pos non hom any order} and \ref{thm: pos non hom 8}.

\section{Preliminaries}
\subsection{Linear Recurrence Sequences}
% A sequence of real numbers \(\mathbf{u} = \tup{u}_{n=1}^\infty\) is
% said to satisfy a \emph{linear recurrence} of order \(k\) if there are
% real numbers \(a_1,\ldots,a_{k+1}\) such that
%\begin{gather}
%  u_{n+k} - a_1 u_{n+k-1} - \cdots - a_{k-1} u_{n+1} - a_k u_n =
%  a_{k+1}
%\label{eq:RECUR}
%\end{gather}
%for all \(n\in \mathbb{N}\).  Such a recurrence is said to be
%\emph{homogeneous} if \(a_{k+1}=0\) and \emph{inhomogeneous} if
%\(a_{k+1}\neq 0\); in the absence of a qualifier, a given sequence is
%assumed to be homogeneous.  The \emph{characteristic polynomial} of
%the recurrence is
%\(p(x) := x^k - a_1x^{k-1} - \cdots - a_{k-1}x - a_k\).  The roots of
%\(p\) are called the \emph{characteristic roots}.  A characteristic root
%of maximum modulus is said to be \emph{dominant} and its modulus is
%the \emph{dominant modulus}. The \emph{multiplicity} of a
%characteristic root \(\gamma\) is the maximal \(m\in \NN\) such that
%\((x-\gamma)^m\) divides \(p(x)\).

Let us start by reformulating the notion of linear recurrence more
abstractly as follows.  Define the \emph{shift operator}
\(E : \RR^{\NN} \rightarrow \RR^{\NN}\) by \(E(f)(n) = f(n+1)\) for a
sequence \(f\in \RR^{\NN}\).  The polynomial ring \(\RR[E]\) acts on the
set of sequences \(\RR^{\NN}\) on the left in a natural way, turning
\(\RR^{\NN}\) into a left \(\RR[E]\) module.  Then a sequence
\(\mathbf{u} = \tup{u_n}_{n=1}^\infty\) satisfies the recurrence equation
(\ref{eq:RECUR}) if and only if
\(p(E) \cdot \mathbf{u} = a_{k+1} \cdot \mathbf{1}\), where \(p\) is the
characteristic polynomial of the recurrence and \(\mathbf{1}\) is the
all-ones sequence.

The following homogenization construction is well known.
\begin{proposition}
  \label{prop:homo}
  Let \(\mathbf{u} = \langle u_n \rangle_{n=1}^\infty\) satisfy an
  inhomogeneous linear recurrence of order \(k\).  Then \(\mathbf{u}\)
  satisfies a homogeneous recurrence of order \(k+1\).
\end{proposition}
\begin{proof}
  By assumption we have that \(p(E) \cdot \mathbf{u} = \mathbf{c}\) for
  some monic polynomial \(p(x)\) of degree \(k\) and constant sequence
  \(\mathbf{c}\).  Writing \(q(x)=(x-1)p(x)\), we have \(q(E) \cdot                                                                                           
  \mathbf{u} = (E-1) \cdot \boldsymbol{c} = 0\).
\end{proof}

We have the following partial converse to Proposition~\ref{prop:homo}.
\begin{proposition}
  \label{prop:inhomo}
  Let \(\mathbf{u} = \langle u_n \rangle_{n=1}^\infty\) satisfy a
  homogeneous linear recurrence of order \(k+1\) with a positive real
  characteristic root \(\rho\).  Then the sequence
  \(\mathbf{v} = \langle v_n \rangle_{n=1}^\infty\) defined by
  \(v_n = \frac{u_n}{\rho^n}\) satisfies an inhomogeneous linear
  recurrence of order \(k\).

\end{proposition}
\begin{proof}
  By assumption, \(\mathbf{u}\) satisfies the recurrence
  \(f(E) \cdot \mathbf{u}=0\) for some monic polynomial
  \(f(x) \in \mathbb{R}[x]\) of degree \(k+1\) that has a positive real
  root \(\rho\).  Define a sequence
  \(\mathbf{v} = \langle v_n \rangle_{n=1}^\infty\) by
  \(v_n := \frac{u_n}{\rho^n}\) for all \(n\in\mathbb{N}\).  Then
  \(\mathbf{v}\) satisfies the recurrence \(g(E) \cdot \mathbf{v} = 0\)
  where \(g\) is the monic polynomial \(g(x) = \rho^{-(k+1)} f(\rho x)\).

But \(g(1)=0\) and hence we have the factorization \(g(x)=(x-1)h(x)\) for
  some monic polynomial \(h(x)\in \mathbb{R}[x]\).  It follows that
  \((E-1)h(E) \cdot \mathbf{v} = 0\) and hence \(h(E) \cdot                                                                                                         
  \mathbf{v}\) is constant, i.e., \(\mathbf{v}\) satisfies an
  inhomogeneous recurrence of order \(k\).
\end{proof}

Let \(\norm{\mathbf{u}}\) denote the binary representation
length\footnote{in general, we denote by \(\norm{\cdot}\) the
  binary-representation length of objects.} of \(\mathbf{u}\).  We
remark that the transformations back and forth between homogeneous and
inhomogeneous LRS can be carried out in polynomial time in \(\norm{u}\)
if the given LRS have real algebraic coefficients.  For an
inhomogeneous LRS \(\mathbf{u}\) of order \(k\), we refer to the
corresponding homogeneous LRS obtained as per
Proposition~\ref{prop:homo} as the \emph{homogenization} of
\(\mathbf{u}\), denoted \(\homLRS(\mathbf{u})\).  The proof of
Proposition~\ref{prop:homo} gives us the following useful property.
\begin{property}
  \label{prop:homogenization eigenvalues}
  The characteristic roots of \(\homLRS(\mathbf{v})\) are the same as
  those of \(\mathbf{v}\), with the same multiplicities, except for the
  characteristic root \(1\),\ which always occurs in
  \(\homLRS(\mathbf{v})\), and whose multiplicity is \(m+1\), where \(m\) is
  the multiplicity of \(1\) in \(\mathbf{v}\).
\end{property}

Consider an LRS \(\mathbf{u}\) with integer coefficients. Since
the characteristic polynomial \(p\) of an LRS \(\mathbf{u}\) has integer
coefficients, the characteristic roots of \(\mathbf{u}\) comprise
real-algebraic roots \(\set{\rho_1,\ldots,\rho_d}\), and conjugate pairs
of complex-algebraic roots
\(\set{\gamma_1,\conj{\gamma_1},\ldots,\gamma_m,\conj{\gamma_m}}\). There
are now univariate polynomials \(A_1,\ldots,A_d\) with real-algebraic
coefficients and \(C_1,\ldots,C_m\) with complex-algebraic coefficients
such that, for every \(n\ge 0\),
\begin{equation*}
  u_n=\sum_{i=1}^{d}A_i(n)\rho_i^n+\sum_{j=1}^m (C_j(n)\gamma_j^n+\conj{C_j(n)}\conj{\gamma}_j^n).
\end{equation*}
This expression is referred to as the \emph{exponential polynomial}
solution of \(\mathbf{u}\). The degree of each of the polynomials is
strictly smaller than the multiplicity of the corresponding root. For
a fixed order \(k\), the coefficients appearing in the polynomials can
be computed in time polynomial in \(\norm{\mathbf{u}}\).

% An LRS is \emph{non-degenerate} if the quotient of every pair of
% distinct roots is not a root of unity. The study of LRS can be
% effectively reduced to that of non-degenerate LRS, by partitioning
% the original LRS to finitely many non-degenerate subsequences
% (see~\ref{Everest RS, OW14b} for details).

We now turn to present two results regarding the asymptotic analysis of LRS. 

The following result due to Braverman~\cite{bra06} enables us to
reason about the complex part of the exponential polynomial above.
\begin{lemma}[Complex Units Lemma]
	\label{lem:braverman}
	Let
        \( \zeta_1, \zeta_2, \ldots, \zeta_m \in
        \mathbf{S}_1\setminus\{1\}\) be distinct complex numbers (where
        \(\mathbf{S}_1=\set{z\in \CC:|z|=1}\)),
	%of modulus \(1\), 
	and let
        \( \alpha_1, \alpha_2, \ldots, \alpha_m \in \CC\setminus\{0\}\).
        Set \(z_n := \sum_{k=1}^m \alpha_k \zeta_k^n\).  Then there
        exists \(c < 0\) such that for infinitely many n,
        \(Re\left(z_n\right) < c\).
\end{lemma}
In particular, Lemma~\ref{lem:braverman} immediately implies that an
LRS without a real dominant characteristic root, is neither positive,
ultimately positive, nor divergent.

% The following theorem, due to Everest, van der Poorten and
% Schlickewei~\cite{EPIW03}, gives a lower bound on the rate of
% absolute divergence.
%\begin{theorem}
%	\label{thm: S units}
%	For any algebraic non-degenerate LRS \(\tup{u_n}\) with dominant
% modulus \(\rho\), and for any \(\epsilon>0\), there exist \(N\in \NN\)
% such \(|u_n|\ge \rho^{(1-\epsilon)n}\) for every \(n>N\).
%\end{theorem}

Finally, the following proposition from~\cite{OW14b} allows us to
bound the growth rate of the low-order terms in the exponential
polynomial of an LRS.
\begin{proposition}
	\label{prop: growth r(n)}
	Consider an LRS \(\mathbf{u}=\tup{u_n}_{n=1}^\infty\) of bounded
        order, with dominant modulus \(\rho\), and write
        \begin{equation*}
          \frac{u_n}{\rho^n} = A(n) + \sum_{i=1}^m \left(
            C_i(n)\lambda_i^n + \conj{C_i(n)}\conj{\lambda}_i^n \right)
            + r(n),
        \end{equation*}
        where \(A\) is a real polynomial, \(C_i\) are non-zero
        complex polynomials, \(\rho\lambda_i\) and
        \(\rho\conj{\lambda}_i\) are conjugate pairs of non-real
        dominant roots of \(\mathbf{u}\), and \(r\) is an exponentially
        decaying function.
	
	We can compute in polynomial time \(\epsilon\in (0,1)\) and \(N\in \NN\) such that
	\begin{align*}
	&\frac1\epsilon=2^{\norm{\mathbf{u}}^{O(1)}} \, ,\\
	&N=2^{\norm{\mathbf{u}}^{O(1)}} \, ,\\
	&\text{for all } n>N, |r(n)|<(1-\epsilon)^n \, .
	\end{align*}
\end{proposition}

\subsection{Mathematical Tools}
In this section we introduce several tools that will be used
throughout the paper.  \subparagraph*{Algebraic Numbers} A complex
number \(\alpha\) is \emph{algebraic} if it is a root of a
polynomial \(p\in \ZZ[x]\). The \emph{defining polynomial} of
\(\alpha\), denoted \(p_\alpha\), is the unique polynomial of the
least degree that vanishes at \(\alpha\), and whose coefficients do
not have common factors other than \(\pm 1\). The
\emph{degree} and the \emph{height} of \(\alpha\) are the degree and the
height (i.e., maximum absolute value of the coefficients) of
\(p_\alpha\), respectively. An algebraic number \(\alpha\) can be
represented by a polynomial that has \(\alpha\) as a root, along with an
approximation of \(\alpha\) by a complex number with rational real and
imaginary parts. We denote by \(\norm{\alpha}\) the representation
length of \(\alpha\).  Basic arithmetic operations as well as equality
testing and comparisons for algebraic numbers can be carried out in
polynomial time (see \cite{CRB13,cohen2013course} for efficient
algorithms).

The following lemma from~\cite{OW14a} is a consequence of the
celebrated lower bound for linear forms in logarithms due to Baker and
W\"ustholz~\cite{baker1993logarithmic}.
% The following celebrated result is due to Baker and W\"ustholz.
%\begin{theorem}[Baker and W\"{u}stholz]
%	\label{thm: baker}
%	Let \(\alpha_1, \ldots, \alpha_m \in \CC\) be algebraic numbers
% different from \(0\) or \(1\), and let \(b_1, \ldots, b_m \in \ZZ\) be
% integers.  Write
%	\[\Lambda = b_1 \log \alpha_1 + \ldots + b_m \log \alpha_m\]
%	Let \(A_1, \ldots, A_m, B \ge e\) be real numbers such that, for
% each \(j \in \lbrace 1, \ldots, m \rbrace\), \(A_j\) is an upper bound
% for the height of \(\alpha_j\), and \(B\) is an upper bound for \(|b_j|\).
% Let \(d\) be the degree of the extension field
% \(\QQ(\alpha_1, \ldots, \alpha_m)\) over \(\QQ\).
%	
%	If \(\Lambda \ne 0\), then
% \(\log |\Lambda| > -(16md)^{2(m+2)} \log A_1 \cdots \log A_m \log B\).
%\end{theorem}
\begin{lemma}
	\label{lem: simple baker}
	There exists \(D \in \NN\) such that, for all algebraic numbers
        \(\lambda, \zeta \in \CC\) of modulus \(1\), and for all
        \(n \ge 2\), if \(\lambda^n \ne \zeta\), then
        \(\left| \lambda^n - \zeta \right| >
        \frac{1}{n^{(\norm{\lambda} + \norm{\zeta})^D}}\).
\end{lemma}

\subparagraph*{Multiplicative Relations} Multiplicative relations
between characteristic roots of an LRS play a key role in our
analysis. The following result, due to Masser~\cite{masser1988linear}
enables us to efficiently elicit these relationships.
\begin{theorem}
	\label{thm: basis mult relations}
	Let \(m\) be fixed, and let \(\lambda_1,\ldots,\lambda_m\) be
        complex algebraic numbers of modulus \(1\). Let
        \(L=\set{(v_1,\ldots,v_m)\in \ZZ^m: \lambda_1^{v_1}\cdots
          \lambda_m^{v_m}=1}\) be the group of multiplicative relations
        between the \(\lambda_i\).  \(L\) has a basis
        \(\set{\Bell_1,\ldots,\Bell_p}\subseteq
        \ZZ^m\) (with \(p\le m\)), where the entries of each of the
        \(\Bell_j\) are all polynomially bounded in
        \(\norm{\lambda_1}+\ldots+\norm{\lambda_m}\). Moreover, such a
        basis can be computed in time polynomial in
        \(\norm{\lambda_1}+\ldots+\norm{\lambda_m}\).
\end{theorem}

\subparagraph*{The First-Order Theory of the Reals} A sentence in the
first-order theory of the reals is of the form
\(Q_1x_1\cdots Q_mx_m \phi(x_1,\ldots,x_m)\) where each \(Q_i\) is a
quantifier (\(\exists\) or \(\forall\)), each \(x_i\) is a real valued
variable, and \(\phi\) is a boolean combination of atomic predicates of
the form \(p(x_1,\ldots,x_m)\sim 0\) for some \(p\in \ZZ[x_1,\ldots,x_m]\)
and \({\sim}\in \set{>,=}\). The first-order theory of the reals admits
quantifier elimination, a famous result due to
Tarski~\cite{tarski1951decision}, whose procedure unfortunately has
non-elementary complexity. In this paper we consider only the case
where the number of variables is uniformly bounded. Then we can
invoke the following result due to
Renegar~\cite{renegar1992computational}.
\begin{theorem}[Renegar]
	\label{thm: renegar}
	Let \(M \in \NN\) be fixed.  Let \(\tau(\boldsymbol y)\) be a formula
        of the first-order theory of the reals.  Assume that the
        number of (free and bound) variables in \(\tau(\boldsymbol y)\) is
        bounded by \(M\).  Denote the degree of \(\tau(\boldsymbol y)\) by
        \(d\) and the number of atomic predicates in \(\tau(\boldsymbol y)\)
        by \(n\).
	
	There is a polynomial time (polynomial in
        \(\norm{\tau(\boldsymbol y)}\)) procedure which computes an
        equivalent quantifier-free formula
	\begin{eqnarray*}
		\chi(\boldsymbol y) = \bigvee_{i=1}^I \bigwedge_{j=1}^{J_i} h_{i,j}(\boldsymbol y) \sim_{i,j} 0,
	\end{eqnarray*}
	where each \(\sim_{i,j}\) is either \(>\) or \(=\), with the following properties:
	\begin{enumerate}
		\item Each of \(I\) and \(J_i\) (for \(1\le i\le I\)) is bounded by \((n+d)^{O(1)}\).
		\item The degree of \(\chi(\boldsymbol y)\) is bounded by \((n+d)^{O(1)}\).
		\item The height of \(\chi(\boldsymbol y)\) is bounded by \(2^{\norm{\tau(\boldsymbol y)}(n+d)^{O(1)}}\).
	\end{enumerate}
\end{theorem}

\subparagraph*{Asymptotic Analysis}

We conclude this section with the following simple lemma
from~\cite{OW14a}.
\begin{proposition}
	\label{prop: bound inverse poly}
	Let \(a\ge 2\) and \(\epsilon\in (0,1)\) be real numbers. Let
        \(B\in \ZZ[x]\) have degree at most \(a^{D_1}\) and height at most
        \(2^{a^{D_2}}\), and assume that \(1/\epsilon\le 2^{a^{D_3}}\) for
        some \(D_1,D_2,D_3\in \NN\). Then there is \(D_4\in \NN\)
        depending only on \(D_1,D_2,D_3\) such that for all
        \(n\ge 2^{a^{D_4}}\), \(\frac{1}{B(n)}>(1-\epsilon)^n\).
\end{proposition}

\section{Divergence}
\label{sec: divergence}
Recall from Theorem~\ref{thm: S units} that an LRS \(\mathbf{u}\) with
dominant modulus \(\rho\) necessarily diverges in absolute value if
\(\rho>1\).  More precisely, if \(\rho>1\) then given \(\varepsilon>0\)
there exists a threshold \(N\) such that \(|u_n|>\rho^{(1-\varepsilon)n}\)
for all \(n>N\).  However this result is \emph{ineffective}---it is not
known how to compute \(N\) given \(\mathbf{u}\) and \(\varepsilon\).

In this section we derive \emph{effective} divergence bounds for
sequences that diverge to \(\infty\) (i.e., sequences that both diverge
in absolute value and that are ultimately positive).  The bounds on
divergence have the following form: for a divergent sequence
\(\mathbf{u}\) with dominant modulus \(\rho=1\) we aim to show that for
every \(n>N\),\ \(u_n>an^d\) for effective constants \(a>0,d\in \NN\), and
\(N\in\NN\).  In case of a dominant modulus \(\rho>1\) we aim to show that
for every \(n>N\),\ \(u_n > \frac{a\rho^n}{n^d}\) for effective constants
\(a>0,d\in \NN\), and \(N\in\NN\).  Henceforth we refer to bounds of these
respective forms as \emph{divergence bounds}.\footnote{Note that not
  only do we seek effective divergence bounds, but also that these
  bounds are asymptotically tighter than the bounds from
  Theorem~\ref{thm: S units} since for any fixed \(d>0\), it is clear
  that \(a\rho^n/n^d\) eventually dominates \(\rho^{(1-\varepsilon)n}\)
  for any \(\varepsilon>0\).}

% In this section we consider the effective divergence problem. As we
% show in the following, when an LRS diverges, the rate of divergence
% depends on the dominant modulus \(\rho\). Specifically, if \(\rho<1\),
% the sequence does not diverge. If \(\rho>1\), the sequence diverges
% roughly at the rate of \(\rho^n\), and if \(\rho=1\) the sequences may
% diverge only if the multiplicity of the characteristic root 1 is at
% least two, in which case the rate of divergence is polynomial.

% More precisely, we formulate the functional version of the
% divergence problem, dubbed \emph{effective divergence} as follows:
% given an LRS \(\mathbf{u}\) with dominant modulus \(\rho\), decide
% whether \(\mathbf{u}\) diverges, and if so, given some rational
% \(\epsilon>0\), if \(\rho>1\), compute \(N\in \NN\) such that for every
% \(n>N\) it holds that \(u_n>\rho^{(1-\epsilon)n}\), and if \(\rho=1\),
% compute \(N,d\in \NN\) and \(a\in \QQ\) such that for every \(n>N\) it
% holds that \(u_n>a n^d\).

In Section~\ref{subsec: div is solvable}, we show how to compute
effective divergence bounds of LRS up to certain orders. Then in
Section~\ref{subsec: div hardness}, we provide hardness results for
the decidability of divergence.

\subsection{Effective Divergence is Solvable}
\label{subsec: div is solvable}
In this section we prove the following theorems:
\begin{theorem}
\label{div-1}
There is a polynomial-time procedure that 
given a rational LRS of order at most 5 decides whether it diverges
and, in case of divergence, outputs divergence bounds.
\label{thm:divergence 5}
\end{theorem}

\begin{theorem}
\label{div-2}
There is a polynomial-time procedure that, given a simple rational LRS
of order at most 8, decides whether it diverges and, in case of
divergence, outputs divergence bounds.
\label{thm:divergence simple 8}
\end{theorem}

The proofs of Theorems~\ref{thm:divergence 5} and~\ref{thm:divergence
  simple 8} build on techniques developed in~\cite{OW14a,OW14b,OW14c},
using a fine-grained analysis in the results thereof, along with some
new ideas. To avoid unnecessary repetition, we sketch the main ideas
of the proofs simultaneously.

Consider an LRS \(\mathbf{u}\) of order \(k\). For uniformity, if
\(\mathbf{u}\) is inhomogeneous, we homogenize it as per
Proposition~\ref{prop:homo}. Thus, either \(k\le 6\) or \(\mathbf{u}\) is
simple and \(k\le 9\), where if \(k=6\) or if \(\mathbf{u}\) is simple and
\(k=9\), then \(\mathbf{u}\) has a special structure according to
Property~\ref{prop:homogenization eigenvalues}.

 %, and let \(\epsilon>0\) be a rational number.

As mentioned in Section~\ref{sec:intro}, we can assume without loss of
generality that \(\mathbf{u}\) is non-degenerate.
% , as we may decompose a degenerate sequence and recast analysis at
% lower orders.
Let \(\rho\) be the dominant modulus of \(\mathbf{u}\), we also note that
if \(\rho<1\), then \(|u_n|\to 0\) as \(n\to \infty\), and in particular the
sequence does not diverge. Thus, we may assume \(\rho\ge 1\).  In
addition, by Lemma~\ref{lem:braverman}, if \(\mathbf{u}\) does not have
a real positive dominant root, then \(u_n \not \rightarrow
\infty\). Thus, we may assume
a real dominant characteristic
root \(\rho>1\).  Note that all other dominant roots must be complex, and
come in conjugate pairs, since if \(-\rho\) were a root, then
\(\mathbf{u}\) would be degenerate.

Writing \(u_n\) as an exponential polynomial and dividing by \(\rho^n\),
we have
\begin{equation}
\label{eq: normalized exp poly}
\frac{u_n}{\rho^n} = A(n) + \sum_{i=1}^m \left( C_i(n)\lambda_i^n + \conj{C_i(n)}\conj{\lambda}_i^n \right) + r(n),
\end{equation}
where \(A\) is a real polynomial, \(C_i\) are non-zero complex
polynomials, \(\rho\lambda_i\) and \(\rho\conj{\lambda}_i\) are conjugate
pairs of non-real dominant characteristic roots of \(\mathbf{u}\) (so
\(|\lambda_i|=1\)), and \(r(n)\) is an exponentially decaying function
(possibly identically zero). More precisely, the degree of each of
\(A(n),C_1(n),\ldots,C_m(n)\) is strictly smaller than the multiplicity
of the corresponding characteristic root.  We can assume that either
\(A(n)\not\equiv 0\) or \(m\neq 0\). Indeed, otherwise we can consider the
LRS \(\tup{\rho^n r(n)}_{n=1}^\infty\), which is of lower order than
\(\mathbf{u}\).

In the following, if \(A(n)\) (resp. \(C_i(n)\) for some \(1\le i\le m\)) is
a constant, we denote it by \(A\) (resp. \(C_i\)).

We proceed to decide divergence by a case analysis of
Equation~\eqref{eq: normalized exp poly}.  \newcounter{divcases}

\refstepcounter{divcases}
\paragraph*{Case \arabic{divcases}: \(\rho=1\) and \(A(n)=A\) is a constant}
\label{case: rho =1 A const}
Note that in this case, \(\frac{u_n}{\rho^n}=u_n\).  Since \(A\) is a
constant, then it does not affect the divergence of \(\mathbf{u}\). We
claim that \(u_n\not\to \infty\). Indeed, by Lemma~\ref{lem:braverman},
the expression
\(\sum_{i=1}^m \left( C_i(n) \lambda_i^n +
  \conj{C_i(n)}\conj{\lambda}_i^n \right)\) becomes negative infinitely
often (regardless of whether \(C_i(n)\) are constants or polynomials),
whereas the effect of \(r(n)\) is exponentially decreasing. Thus,
\(\mathbf{u}\) does not diverge.

\refstepcounter{divcases}
\paragraph*{Case \arabic{divcases}: \(\rho=1\), \(A(n)\) is not a constant, and every \(C_i\) is a constant}
\label{case: rho=1 A not const every C const}
% The case where \(A(n)\) is not a constant can arise in two
% scenarios. Either \(k\le 6\), or, by Property~\ref{prop:homogenization
% eigenvalues}, \(k\le 9\) and \(\mathbf{u}=\homLRS(\mathbf{v})\), where
% \(\mathbf{v}\) is a simple LRS of order at most \(8\) whose dominant
% characteristic root is of modulus \(1\). We start by addressing the
% latter scenario. Note that since \(\mathbf{v}\) is simple, then the
% multiplicity of the characteristic root \(1\) in \(\mathbf{u}\) is at
% most \(2\), so \(A(n)\) is linear. Moreover, all other characteristic
% roots of \(\mathbf{u}\) have multiplicity \(1\).
In this case we can rewrite Equation~\eqref{eq: normalized exp poly} as 
\begin{equation}
\label{eq: exp poly root 1 simple}
u_n = A(n) + \sum_{i=1}^m \left( C_i \lambda_i^n + \conj{C_i}\conj{\lambda}_i^n \right) + r(n).
\end{equation}
Since \(|\lambda_i|=1\) for all \(i\), and since \(r(n)\) is exponentially
decreasing, then clearly \(u_n\to \infty\) iff the leading coefficient
of \(A(n)\) is positive.

Recall that since \(\rho=1\), then if \(\mathbf{u}\) diverges, there exist
\(N,d\in \NN\) and \(a>0\) such that \(u_n\ge a n^d\) for all \(n>N\). We now
show how to effectively compute \(N\), \(d\), and \(a\).

From Proposition~\ref{prop: growth r(n)}, we can compute in polynomial
time \(\epsilon\in (0,1)\) and \(N_1\in \NN\) such that
\(r(n)<(1-\epsilon)^n<1\) for all \(n>N_1\). We thus have that
\(u_n\ge A(n)-2\sum_{i=1}^m|C_i|-1\), and we can easily compute
\(N_2\in \NN\) and \(a\in \QQ\) (depending on the coefficients of \(A(n)\))
such that for all \(n>N_2\) we have \(A(n)-2\sum_{i=1}^m|C_i|-1\ge an^d\),
where \(d\) is the degree of \(A(n)\). Taking \(N=\max\set{N_1,N_2}\), we
conclude this case.

\refstepcounter{divcases}
\paragraph*{Case \arabic{divcases}: \(\rho=1\), \(A(n)\) is not a constant, and there exists a non-constant \(C_i(n)\)}
\label{case: rho=1 A not const exists C non const}
We notice that if there exists a non-constant \(C_i(n)\), it follows by
Property~\ref{prop:homogenization eigenvalues} that \(\mathbf{u}\) is
not obtained by homogenizing a simple LRS. That is, we are in the case
where \(k\le 6\). In the notations of Equation~\eqref{eq: normalized exp
  poly}, we then have that \(m=1\), \(A(n)\) is linear, \(C_1(n)\) is
linear, and \(r(n)\equiv 0\). Indeed, this corresponds to the case where
the characteristic roots of \(u_n\) are \(1,\lambda,\conj{\lambda}\), each
with multiplicity \(2\). Let \(A(n)=a_1 n+ b_1\) and \(C_1(n)=a_2 n+b_2\),
then we can write
\begin{equation*}
  u_n=a_1 n+ b_1+ (a_2 n+ b_2)\lambda^n+ (\conj{a_2} n +
  \conj{b_2})\conj{\lambda}^n=n(a_1+a_2\lambda^n +
  \conj{a_2}\conj{\lambda^n})+(b_1+b_2\lambda^n +
  \conj{b_2}\conj{\lambda^n})
\end{equation*}
Since \(|(b_1+b_2\lambda^n+\conj{b_2}\conj{\lambda^n})|\) is bounded,
then \(u_n\) diverges iff
\(n(a_1+a_2\lambda^n+\conj{a_2}\conj{\lambda^n})\) diverges. Let
\(\theta=\arg \lambda\) and \(\phi=\arg a_2\). We have
\(n(a_1+a_2\lambda^n+\conj{a_2}\conj{\lambda^n})=n(a_1+2|a_2|\cos(n\theta+\phi))\).

Observe that since \(\mathbf{u}\) is non-degenerate, then \(\theta\) is
not a rational multiple of \(\pi\). It follows that
\(\set{[n\theta+\phi]_{2\pi}:n\in \NN}\) (where \([x]_{2\pi}=x-2\pi j\)
where \(j\) is the unique integer such that \(0\le x-2\pi j<2\pi\)) is
dense in \([0,2\pi)\), so \(\set{\cos(n\theta+\phi):n\in \NN}\) is dense
in \([-1,1]\).  Again, we split into cases.

\(\bullet\) If \(a_1>2|a_2|\), we have that \(u_n\) diverges. Then we can
compute in polynomial time a rational \(\epsilon>0\) and \(N\in \NN\) such
that \(a_1-2|a_2|>\epsilon\) and \(n(a_1+2|a_2|)-(b_1-2|b_2|)>\epsilon n\)
for all \(n>N\). We then have that \(u_n>\epsilon n\) for all \(n>N\), thus
concluding effective decidability of divergence in this case.

\(\bullet\) If \(a_1<2|a_2|\), then \(u_n\) does not diverge, as it becomes
negative infinitely often, by the density argument above.

\(\bullet\) The remaining case is when \(a_1=2|a_2|\), and the expression
above becomes \(na_1(1+\cos(n\theta+\phi))\). We show that in this case,
\(u_n\) does not diverge.

By Taylor approximation, for every \(x\in (-\pi,\pi]\) it holds that \(1-\cos(x)\le \frac{x^2}{2}\). 
For \(n\in \NN\), write \(\Lambda(n)=n\theta+\phi-(2j+1)\pi\), where \(j\in \ZZ\) is the unique integer such that \(-\pi<\Lambda(n)\le \pi\). 
We now have that
\begin{align*}
na_1(1+\cos(n\theta+\phi)) = na_1(1-\cos(n\theta+\phi+\pi))=
  na_1(1-\cos(\Lambda(n)))<na_1\frac{\Lambda(n)^2}{2} \, .
\end{align*}
By Dirichlet's Approximation Theorem, we have that
\(|\Lambda(n)|<\frac{t}{n}\) for infinitely many values of \(n\), where
\(t\) is a constant depending on \(\phi\). Thus, we have
\(na_1\frac{\Lambda(n)^2}{2}<\frac{a_1t^2}{2n}\) for infinitely many
values of \(n\).  It follows that \(u_n\) is infinitely often bounded by a
constant, so it does not diverge.

\refstepcounter{divcases}
\paragraph*{Case \arabic{divcases}: \(\rho>1\) and there exists a non-constant \(C_i(n)\)}
\label{case: rho>1 exists non const C}

As in Case~\ref{case: rho=1 A not const exists C non const}, it holds
that \(k\le 6\). Moreover, since \(\rho>1\), then whether or not
\(\mathbf{u}\) was obtained by homogenization, the characteristic root
\(1\) (if it exists) is captured in \(r(n)\). Therefore, we have that
\(m=1\), \(C_1\) is linear, and \(A(n)=A\) is constant.  Let \(C_1\) have
leading coefficient \(b \ne 0\). By Lemma~\ref{lem:braverman}, there
exists \(\epsilon > 0\) such that
\(b\lambda^n + \conj{b}\conj{\lambda}^n < -\epsilon\) infinitely often.
Then \(C_1(n) \lambda_1^n + \conj{C_1(n)} \conj{\lambda}_1^n\) (and
hence \(u_n\)) is unbounded below, so \(u_n\) does not diverge.

\refstepcounter{divcases}
\paragraph*{Case \arabic{divcases}: \(\rho>1\), \(A(n)\) is not a constant, and every \(C_i\) is a constant}
\label{case: rho>1 A non const every C const}
Since \(A(n)\) is not a constant and \(\rho>1\), this case may only arise
for \(k\le 6\) and \(m\le 1\). We write
\begin{equation*}
\frac{u_n}{\rho^n} = A(n) + C_1\lambda_1^n + \conj{C_1}\conj{\lambda}_1^n + r(n).
\end{equation*}
where if \(m=0\) then take \(C_1=0\).

If \(A(n)\) has a negative leading coefficient, then \(u_n\) is unbounded
from below, and in particular \(u_n\) does not diverge.

If \(A(n)\) has a positive leading coefficient, we can compute in
polynomial time \(N_0 \in \NN\) and a rational \(\epsilon_0>0\) such that
\(A(n) - 2|C_1|>2\epsilon_0\) for all \(n>N_0\). By Proposition~\ref{prop:
  growth r(n)}, we can also compute in polynomial time \(N_1\in \NN\)
and \(\epsilon_1\in (0,1)\) such that \(|r(n)|<(1-\epsilon_1)^n\) for all
\(n>N_1\). Taking \(N_2\ge \log_{1-\epsilon_1}\epsilon_0\), we have that
for all \(n>\max\set{N_0,N_1,N_2}\), \(|r(n)|< \epsilon_0\), and thus
\begin{equation*}
  \frac{u_n}{\rho^n} \ge  A(n) -2|C_1| +r(n)
  \ge A(n) -2|C_1|-\epsilon_0
  > 2\epsilon_0-\epsilon_0=\epsilon_0\, .
\end{equation*}
Thus we have \(u_n \geq \epsilon_0 \rho^n\) for all \(n > \max\set{N_0,N_1,N_2}\), which immediately yields 
effective divergence bounds in this case.

\refstepcounter{divcases}
\paragraph*{Case \arabic{divcases}: \(\rho>1\), \(A(n)=A\) is a constant, and every \(C_i\) is a constant}
\label{case: rho>1 A const every C const}
This case is the most involved, and utilizes deep mathematical results. Our proof works along the lines of~\cite{OW14c}. For completeness, the full proof can be found in Appendix~\ref{app: proofs divergence}.

We rewrite Equation~\eqref{eq: normalized exp poly} as
\begin{equation}
\label{eq: normalized exp poly constants}
\frac{u_n}{\rho^n} = A + \sum_{i=1}^m \left( C_i\lambda_i^n + \conj{C_i}\conj{\lambda}_i^n \right) + r(n).
\end{equation}

Observe that \(m\le 3\). Indeed, if \(k\le 8\) this is trivial, and if \(k=9\) then by Property~\ref{prop:homogenization eigenvalues}, \(1\) must be a non-dominant characteristic root of \(\mathbf{u}\), so \(r(n)\not\equiv 0\) and thus \(m\le 3\).

In the following, we handle the case \(m=3\). The cases where \(m<3\) are similar and slightly simpler.

Let \(L=\set{(v_1,\ldots,v_3)\in \ZZ^3: \lambda_1^{v_1}\cdots \lambda_3^{v_3}=1}\), and let \(\set{\Bell_1, \ldots , \Bell_p }\) be a basis for \(L\) of cardinality \(p\). Write \(\Bell_{q}=(\ell_{q,1},\ldots,\ell_{q,3})\) for \(1\le q\le p\). From Theorem~\ref{thm: basis mult relations}, such a basis can be computed in polynomial time, and moreover, each \(\ell_{q,j}\) may be assumed to have magnitude polynomial in \(\norm{u}\).

Consider the set \(\TT=\{(z_1,z_2,z_3)\in \CC^3: |z_1|=|z_2|=|z_3|=1 \text{ and for each } 1\le q\le p,\) \( z_1^{\ell_{q,1}}z_2^{\ell_{q,2}}z_3^{\ell_{q,3}}=1\}\).

Define \(h:\TT\to \RR\) by setting \(h(z_1,z_2,z_3)=\sum_{i=1}^3 (C_iz_i+\conj{C_i}\conj{z_i})\), so that for every \(n\in \NN\),
\(\frac{u_n}{\rho^n}=A+h(\lambda_1^n,\lambda_2^n,\lambda_3^n)+r(n)\). 
By Kronecker's theorem on inhomogeneous Diophantine approximation~\cite{cassels1965introduction}, the set \(\set{\lambda_1^n,\lambda_2^n,\lambda_3^n:n\in \NN}\) is a dense subset of \(\TT\). Since \(h\) is continuous, it follows that \(\inf\set{h(\lambda_1^n,\lambda_2^n,\lambda_3^n): n\in \NN}=\min h|_\TT=\mu\) for some \(\mu\in \RR\). 

%We now claim that \(\mu\) is an algebraic number, computable in polynomial time, with \(\norm{\mu}=\norm{\mathbf{u}}^{O(1)}\).
%We can represent \(\mu\) via the following formula \(\tau(y)\):
%\begin{equation*}\exists (\zeta_1,\zeta_2,\zeta_3)\in \TT [h(\zeta_1,\zeta_2,\zeta_3)=y \wedge \forall (z_1,z_2,z_3)\in \TT, y\le h(z_1,z_2,z_3)].\end{equation*} 
%Note that \(\tau(y)\) is not a formula in the first-order theory of the reals, as it involves complex numbers. However, we can rewrite it as a sentence in the first-order theory of the reals by representing the real and imaginary parts of each complex quantity and combining them using real arithmetic (see~\cite[ICALP] for details). In addition, the obtained formula \(\tau'(y)\) is of size polynomial in \(\norm{\mathbf{u}}\). By Theorem~\ref{thm: renegar}, we can then compute in polynomial time an equivalent quantifier-free formula
%\begin{equation*}\chi(x)=\bigvee_{i=1}^I\bigwedge_{j=1}^{J_i}h_{i,j}\sim_{i,j}0.\end{equation*}
%Recall that each \(\sim_{i,j}\) is either \(>\) or \(=\). Now \(\chi(x)\) must have a satisfiable disjunct, and since the satisfying assignment to \(y\) is unique (namely \(y=\mu\)), this disjunct must comprise at least one equality predicate. Since Theorem~\ref{thm: renegar} guarantees that the degree and height of each \(h_{i,j}\) are bounded by \(\norm{\mathbf{u}}^{O(1)}\) and \(2^{\norm{\mathbf{u}}^{O(1)}}\) respectively, we immediately conclude that \(\mu\) is
%an algebraic number and with \(\norm{\mu}=\norm{\mathbf{u}}^{O(1)}\).

In the full proof, we show that \(\mu\) is algebraic, computable in polynomial time, with \(\norm{\mu}=\norm{\mathbf{u}}^{O(1)}\).

We now split to cases according to the sign of \(A+\mu\).

\(\bullet\) If \(A+\mu<0\), then \(\mathbf{u}\) is infinitely often negative, and does not diverge.

\(\bullet\) If \(A+\mu>0\), then \(\mathbf{u}\) diverges, and we obtain an effective bound similarly to Case~\ref{case: rho>1 A non const every C const}.

%and it remains to show an effective bound. We can compute in polynomial time a rational \(\epsilon_0>0\) such that \(a + \mu>2\epsilon_0\). By Proposition~\ref{prop: growth r(n)}, we can also compute in polynomial time \(N_1\in \NN\) and \(\epsilon_1\in (0,1)\) such that \(|r(n)|<(1-\epsilon_1)^n\) for all \(n>N_1\). Taking \(N_2\ge \log_{1-\epsilon_1}\epsilon_0\), we have that for all \(n>\max\set{N_1,N_2}\), \(|r(n)|< \epsilon_0\), and thus  \begin{equation*}\frac{u_n}{\rho^n} =  A + h(\lambda_1, \ldots, \lambda_m)+r(n)
%\ge A + \mu-\epsilon_0\\
%> 2\epsilon_0-\epsilon_0=\epsilon_0\end{equation*}
%The constant \(\epsilon_0\) can be thought of as an inverse polynomial for the purpose of applying Lemma~\ref{lem: computable bound LRS}, so we conclude the effective decidability of divergence in this case.

\(\bullet\) It remains to analyze the case where \(A+\mu=0\). To this end, let \(\lambda_j=e^{i\theta_j}\) and \(C_j=|C_j|e^{i\phi_j}\) for \(1\le j\le 3\). From Equation~\eqref{eq: normalized exp poly constants} we have 
\begin{equation*}
  \frac{u_n}{\rho^n}=A+\sum_{j=1}^3 2|C_j|\cos(n\theta_j+\phi_j)+r(n).
\end{equation*}
We further assume that all the \(C_j\) are non-zero. Indeed, if this does not hold, we can recast our analysis in lower dimension.

In the full proof, we use zero-dimensionality results to show 
%We now claim 
that \(h\) achieves its minimum \(\mu\) over \(\TT\) only at a finite set of points \(Z=\set{(\zeta_1,\zeta_2,\zeta_3)\in \TT: h(\zeta_1,\zeta_2,\zeta_3)=\mu}\).

We concentrate on the set \(Z_1\) of first coordinates of \(Z\). Write
\begin{align*}
\tau_1(x)=\exists z_1 (\re(z_1)=x\wedge z_1\in Z_1),\\
\tau_2(y)=\exists z_1 (\im(z_1)=y\wedge z_1\in Z_1).
\end{align*}

By rewriting these formulas in the first-order theory of the reals, we are able to show, using Theorem~\ref{thm: renegar}, that 
%Similarly to our earlier constructions \(\tau_1(x)\) is equivalent to a formula \(t'_1(x)\) in the in the first-order theory of the reals, over a bounded number of real variables, with \(\norm{\tau'_1(x)}=\norm{\mathbf{u}}^{O(1)}\).
%Thanks to Theorem~\ref{thm: renegar}, we then obtain an equivalent quantifier-free formula
%\begin{equation*}\chi_1(x)=\bigvee_{i=1}^I\bigwedge_{j=1}^{J_i}h_{i,j}\sim_{i,j}0.\end{equation*}
%Note that since there can only be finitely many \(\hat{x}\in \RR\) such that \(\chi_1(\hat{x})\) holds, each disjunct of \(\chi_1(\hat{x})\) must comprise at least one equality predicate, or can otherwise be entirely discarded as having no solution. 
%A similar exercise can be carried out with \(\tau_2(x)\). The bounds on the degree and height of each \(h_{i,j}\) in \(\chi_1(x)\) and \(\chi_2(y)\) then enables us to conclude that 
any \(\zeta_1=\hat{x}+i \hat{y}\in Z_1\) is algebraic, and moreover satisfies \(\norm{\zeta_1}= \norm{\mathbf{u}}^{O(1)}\). 
In addition, we show that the cardinality of \(Z_1\) is at most polynomial in \(\norm{\mathbf{u}}\).

Since \(\lambda_1\) is not a root of unity, for each \(\zeta_1\in Z_1\) there is at most one value of \(n\) such that \(\lambda_1^n=\zeta_1\). 
Theorem~\ref{thm: basis mult relations} then entails that this value (if it
exists) is at most \(M = \norm{\mathbf{u}}^{O(1)}\), which we can take to be uniform across all \(\zeta_1\in Z_1\). 
We can now invoke Corollary~\ref{lem: simple baker} to conclude that, for \(n > M\), and for all \(\zeta_1\in Z_1\), we have
\begin{equation}
\label{eq: baker usage zeta simple}
|\lambda_1^n-\zeta_1|>\frac{1}{n^{\norm{\mathbf{u}}^{D}}},
\end{equation}
where \(D\in \NN\) is some absolute constant.

Let \(b>0\) be minimal such that the set
\begin{equation*}
  \set{z_1\in \CC: |z_1|=1 \text{ and, for all } \zeta_1\in Z_1, |z_1-\zeta_1|\ge \frac1b}
\end{equation*} 
is non empty. Thanks to our bounds on the cardinality of \(Z_1\), we can use the first-order theory of the reals, together with Theorem~\ref{thm: renegar}, to conclude that \(b\) is algebraic and \(\norm{b}=\norm{\mathbf{u}}^{O(1)}\).

Define the function \(g:[b,\infty)\to \RR\) as follows:
\begin{align*}
g(x)=\min\{h(z_1,z_2,z_3)-\mu: (z_1,z_2,z_3)\in \TT \text{ and for all } \zeta_1\in Z_1, |z_1-\zeta_1|\ge \frac{1}{x}\}.
\end{align*}
In the full proof we show that we can compute in polynomial time a polynomial 
%It is clear that \(g\) is continuous and \(g(x) > 0\) for all \(x\in [b,\infty)\). 
%Moreover, \(g\) can be translated in polynomial time into a function in the first-order theory of the reals over a bounded number of variables. 
%It follows from Proposition 2.6.2 of~\cite{Real Algebraic Geometry, coste roy bochnak} (invoked with the function \(1/g\)) that there is a polynomial 
\(P\in \ZZ[x] \) such that, for all \(x\in [b,\infty)\),
\begin{equation}
\label{eq: bound on g simple}
g(x)\ge \frac{1}{P(x)}
\end{equation}
%Moreover, an examination of the proof of \cite{Real Algebraic Geometry,Prop. 2.6.2} reveals that \(P\) is obtained through a process which hinges on quantifier elimination. 
%By Theorem~\ref{thm: renegar}, we are therefore able to conclude that 
with \(\norm{P}=\norm{\mathbf{u}}^{O(1)}\)
%, a fact which relies among others on our upper bounds for \(\norm{b}\).

By Proposition~\ref{prop: growth r(n)} we can find \(\epsilon\in (0,1)\) and \(N=2^{\norm{\mathbf{u}}^{O(1)}}\) such that for all \(n>N\), we have \(|r(n)|<(1-\epsilon)^n\), and moreover \(1/\epsilon=2^{\norm{\mathbf{u}}^{O(1)}}\). 
In addition, by Proposition~\ref{prop: bound inverse poly}, there is \(N'=2^{\norm{\mathbf{u}}^{O(1)}}\) such that for every \(n\ge N'\)
\begin{equation}
\label{eq: bound on P simple}
\frac{1}{2P(n^{\norm{\mathbf{u}}^D})}>(1-\epsilon)^n.
\end{equation}
Combining Equations~\eqref{eq: normalized exp poly constants}--\eqref{eq: bound on P simple}, we get
\begin{align*}
\frac{u_n}{\rho^n}&=A+h(\lambda_1^n,\lambda_2^n,\lambda_3^n)+r(n)
\ge -\mu+h(\lambda_1^n,\lambda_2^n,\lambda_3^n)-(1-\epsilon)^n 
\ge g(n^{\norm{\mathbf{u}}^D})-(1-\epsilon)^n\\
&\ge \frac{1}{P(n^{\norm{\mathbf{u}}^D})}-(1-\epsilon)^n
= \frac{1}{2P(n^{\norm{\mathbf{u}}^D})}+\frac{1}{2P(n^{\norm{\mathbf{u}}^D})}-(1-\epsilon)^n
\ge \frac{1}{2P(n^{\norm{\mathbf{u}}^D})}
\end{align*}
provided \(n>\max\set{M,N,N'}\). We thus have that \(\frac{u_n}{\rho^n}\) is eventually lower bounded by an inverse polynomial and hence 
we have effective divergence bounds in this case.

Finally, Cases~\ref{case: rho =1 A const}--\ref{case: rho>1 A const every C const} allow us to conclude both Theorem~\ref{thm:divergence 5} and Theorem~\ref{thm:divergence simple 8}. 
%%%%%%%%%%%%%%%%%

\subsection{Hardness of Divergence}
\label{subsec: div hardness}
We now turn to show lower bounds for the divergence
problem. Surprisingly, our lower bounds hold already for homogeneous
LRS, and for the divergence decision problem, even without requiring
effectively computable bounds.

In~\cite{OW14b}, it is shown that the Ultimate Positivity problem for
homogeneous LRS of order at least 6 is hard, in the sense that if
Ultimate Positivity is decidable for such LRS, then certain hard open
problems in Diophantine approximation become solvable. We show
hardness of divergence for homogeneous LRS of order at least 6 by
reducing from Ultimate Positivity.

\begin{theorem}
\label{hardness}
Ultimate Positivity is reducible to Divergence.
\end{theorem}
\begin{proof}
  We show a reduction from the Ultimate Positivity problem for
  non-degenerate LRS of order \(6\), shown to be hard
  in~\cite{OW14a}. The key ingredient in the reduction is
  Theorem~\ref{thm: S units}.
	
  Consider a non-degenerate homogeneous LRS \(\tup{u_n}\) of order \(6\)
  with dominant modulus \(\rho\), and let \(\mu=\max\set{2,\frac2\rho}\),
  then the sequence \(v_n=\mu^n u_n\) is a non-degenerate homogeneous
  LRS of order \(6\) with dominant modulus \(\mu\rho\ge 2\). By
  Theorem~\ref{thm: S units}, taking \(\epsilon=\frac12\), it follows
  that there exists \(N\in \NN\) such that \(|v_n|\ge 2^{n/2}\) for every
  \(n>N\).  It immediately follows that \(v_n\) is ultimately positive iff
  \(v_n\to \infty\). Clearly, however, \(v_n\) and \(u_n\) have the same
  sign, and therefore \(u_n\) is ultimately positive iff \(v_n\) diverges,
  and we are done.
\end{proof}

\section{Positivity and Ultimate Positivity}
\label{sec:pos and upos}

% OW14a - On the Positivity Problem for Simple Linear Recurrence
% Sequences OW14b - Positivity Problems for Low-Order Linear
% Recurrence Sequences. In Proc. SODA OW14c - Ultimate Positivity is
% Decidable for Simple Linear Recurrence Sequences. In Proc. ICALP

In this section we study the Positivity and Ultimate Positivity
problems for inhomogeneous LRS. These problems were studied
in~\cite{OW14b,OW14a, OW14c} for homogeneous LRS. Using
Proposition~\ref{prop:homo} and some careful analysis, we extend the
decidability results to inhomogeneous LRS.

We start by citing some results from~\cite{OW14b, OW14a, OW14c}, split
to upper and lower bounds.
\begin{theorem}[Upper Bounds from~\cite{OW14b,OW14a, OW14c}]
  \label{thm: cite results upper}\
  \begin{enumerate}
  \item\label{itm: pos and upos dec 5} Positivity and Ultimate
    Positivity are decidable for homogeneous LRS of order 5 or less
    with complexities in \coNPposslp and \Ptime, respectively.
  \item\label{itm: pos hom dec 9} Positivity is decidable for simple
    homogeneous LRS of order 9 or less with complexity in \coNPposslp.
  \item\label{itm: upos dec sim any} Ultimate Positivity is decidable
    for simple homogeneous LRS of any order with complexity in \Ptime.
  \item\label{itm: upos eff sim 9} Effective Ultimate Positivity is
    solvable for simple homogeneous LRS of order 9 or less with
    complexity in \Ptime.
  \end{enumerate}
\end{theorem}

The following notion of hardness will be made precise in
Section~\ref{subsec: lower bounds ult pos}.
\begin{theorem}[Lower Bounds from~\cite{OW14b, OW14a, OW14c}]
  \label{thm: cite results lower}
  Positivity and Ultimate Positivity for LRS of order at least 6 are
  hard with respect to certain hard open problems in Diophantine
  approximation.
\end{theorem}

\subsection{Upper Bounds}
We proceed to prove analogous results to Theorem~\ref{thm: cite
  results upper} for inhomogeneous LRS.

Theorem~\ref{thm: cite results upper}(\ref{itm: pos and upos dec 5}.)
along with Proposition~\ref{prop:homo} readily give us the following:
\begin{theorem}
  \label{thm: non hom pos and upos dec 4}
  Positivity and Ultimate Positivity are decidable for inhomogeneous
  LRS of order \(4\) or less, with complexity in \coNPposslp and \Ptime\!\!,
  respectively.
\end{theorem}

For simple LRS, things become more involved, as
Proposition~\ref{prop:homo} does not preserve simplicity. However,
Property~\ref{prop:homogenization eigenvalues} shows that simplicity
is almost preserved, up to the multiplicity of the characteristic root
1. As we now show, this is sufficient to obtain upper bounds for
inhomogeneous simple LRS.

We start by addressing effective Ultimate Positivity, which we then
use for addressing Positivity.

\begin{theorem}
  \label{thm: effective ult pos non hom 8}
  Effective Ultimate Positivity is solvable in polynomial time for
  simple inhomogeneous LRS of order \(8\) or less.
\end{theorem}
\begin{proof}
  Let \(\mathbf{v}\) be a simple, non-degenerate, inhomogeneous LRS or
  order \(8\) or less, and consider the homogeneous LRS
  \(\mathbf{u}=\homLRS(\mathbf{v})\). By Proposition~\ref{prop:homo},
  \(\mathbf{u}\) is of order at most \(9\). If \(\mathbf{u}\) is a simple
  LRS, then by~\cite{OW14a} we can effectively decide its Ultimate
  Positivity. We hence assume that \(\mathbf{u}\) is not simple.
	
  By Property~\ref{prop:homogenization eigenvalues}, it follows that
  the characteristic roots of \(\mathbf{u}\) all have multiplicity \(1\),
  apart from the characteristic root \(1\) which has multiplicity
  \(2\). Consider the dominant modulus \(\rho\) of \(\mathbf{u}\). If
  \(\rho>1\), then by writing the exponential polynomial of
  \(\mathbf{u}\), we have
  \begin{equation}
    \label{eq: exp poly effective ult pos}
    \frac{u_n}{\rho^n}=a+\sum_{i=1}^m (c_i\lambda_i^n+\conj{c_i}\conj{\lambda_i^n})+r(n)
  \end{equation}
  with \(a\in \RR\), \(c_i\in \CC\), and \(|\lambda_i|=1\) for
  every \(1\le i\le m\), and \(|r(n)|\) exponentially decaying.
  Crucially, since \(1\) is not a dominant characteristic root, its
  effect is enveloped in \(r(n)\). Specifically, we observe that the
  analysis of effective Ultimate Positivity in~\cite{OW14a} only
  relies on Proposition~\ref{prop: growth r(n)}. Since this holds in
  the case at hand, we can effectively decide Ultimate Positivity when
  \(1\) is not a dominant characteristic root.
	
  Finally, if \(1\) is a dominant characteristic root, the exponential
  polynomial of \(\mathbf{u}\) can be written as
  \begin{equation}
    \label{eq: exp poly effective ult pos 1 dominant}
    u_n=A(n)+\sum_{i=1}^m (c_i\lambda_i^n+\conj{c_i}\conj{\lambda_i^n})+r(n).
  \end{equation}
  We observe that in this case, \(u_n\) is ultimately positive iff it
  diverges (indeed, clearly \(|u_n|\to \infty\)). Thus, we can reduce
  the problem to divergence, and proceed with the analysis as in
  Section~\ref{sec: divergence} Case~\ref{case: rho=1 A not const
    every C const}.
	
  %	with \(A(n)=an+b\) a linear polynomial, and \(a\neq 0\). Let
  % \(C=\max_{1\le i\le m} \set{|c_i|}\), and Observe that
  % \(|\sum_{i=1}^m (c_i\lambda_i^n+\conj{c_i}\conj{\lambda_i^n})|\le 2
  % m C\le 8C\) (trivially \(m\le 4\)).
  %	
  %	If \(a<0\), then clearly \(u_n\to -\infty\), and is thus not
  % ultimately positive. If \(a>0\), by Proposition~\ref{prop: growth
  % r(n)}, we can compute in polynomial time \(\epsilon\in (0,1)\) and
  % \(N_0\in \NN\) such that \(|r(n)|<(1-\epsilon)^n<1\) for all
  % \(n>N_0\). We can now easily compute \(N_2\in \NN\) and \(q>0\)
  % (depending on the coefficients of \(A(n)\)) such that for all
  % \(n>N_2\) we have \(A(n)-C-1\ge qn>0\). Taking
  % \(N=\max\set{N_0,N_1,N_2}\), we conclude that \(u_n\) is ultimately
  % positive, and we can compute the relevant bound. Observe that by
  % Proposition~\ref{prop: growth r(n)} we have
  % \(N=2^{\mathbf{u}^{O(1)}}\).
  %	
  This concludes the proof that Ultimate Positivity is effectively
  decidable for simple inhomogeneous LRS of order at most \(8\).
\end{proof}

Similarly to Theorem~\ref{thm: effective ult pos non hom 8}, we are
able to conclude the following result, whose proof can be found in
Appendix~\ref{app: proof of ultimate pos non hom any order}.

\begin{theorem}
  \label{thm: ultimate pos non hom any order}
  Ultimate Positivity is decidable in polynomial time for simple
  inhomogeneous LRS of any order.
\end{theorem}

Finally, using Theorem~\ref{thm: effective ult pos non hom 8}, we can
solve the Positivity problem (see Appendix~\ref{app: proof of pos non
  hom 8} for the proof).
\begin{theorem}
  \label{thm: pos non hom 8}
  Positivity is decidable for simple inhomogeneous LRS of order \(8\) or
  less, with complexity in \coNPposslp.
\end{theorem}
% \begin{proof}
%   Given the proof of Theorem~\ref{thm: effective ult pos non hom 8},
%   Positivity is now easily decidable: given an inhomogeneous simple
%   LRS \(\mathbf{u}\) of order at most \(8\), decide if its ultimately
%   positive, and if so - compute the bound from which it is
%   ultimately positive. Then deciding Positivity amounts to checking
%   a finite number of elements.
%	
%   Note that the bound computed in Theorem~\ref{thm: effective ult
%   pos non hom 8} is \(N=2^{\mathbf{u}^O(1)}\). This implies that
%   checking whether an ultimately-positive LRS is \emph{not} positive
%   can be done using a \emph{guess-and-check} procedure, and
%   employing PoSSLP in order to compute double exponential
%   numbers. This yields an \(NP^{PosSLP}\) algorithm. Thanks
%   to~\cite{Allender}, we obtain an upper bound of
%   \(coNP^{PP^{PP^{PP}}}\) for Positivity (see~\cite{OW14a} for
%   details).
%\end{proof}

\subsection{Lower Bounds}
\label{subsec: lower bounds ult pos}
We now turn to study lower bounds, proving analogous results to
Theorem~\ref{thm: cite results lower} for inhomogeneous LRS. Similarly
to~\cite{OW14a}, the hardness results we present are with respect to
long standing open problems in Diophantine approximation.  Before
stating our results, we require some definitions from Diophantine
approximation. We refer the reader
to~\cite{niven2008diophantine,OW14a} for comprehensive references.

For any \(x\in \RR\), we define the \emph{Lagrange constant} of \(x\) as
\begin{equation*}
  L_\infty(x)=\inf\{c\in \RR: |x-\frac{n}{m}|\le \frac{c}{m^2} \text{ for infinitely many }m,n\in \ZZ\},
\end{equation*}
and the \emph{approximation type} of \(x\) as 
\begin{equation*}
  L(x)=\inf\{c\in \RR: |x-\frac{n}{m}|\le \frac{c}{m^2} \text{ for some }m,n\in \ZZ\}.
\end{equation*}
For the vast majority of transcendental numbers, the Lagrange constant
and the approximation type are unknown, despite significant
work~\cite{cusick1989markoff,OW14a}, and the problem of computing them
is a major open problem. In the following, we show that the
decidability of Ultimate Positivity (resp. Positivity) for
inhomogeneous LRS of order 5 or more would imply a major breakthrough
in computing the Lagrange constant (resp. approximation type) for a
large class of transcendental numbers.

\begin{theorem}
  If Ultimate Positivity is decidable for inhomogeneous rational LRS
  of order at least 5 then there is an algorithm that computes the
  Lagrange constant of any number \(\theta/2\pi\) such that
  \(e^{i\theta}\) has rational real and imaginary parts.
  % \footnote{the development in~\cite{OW14a} actually assumes that
  % the real and imaginary parts of \(e^{i\theta}\) are rational
  % numbers. %However this is only due to~\cite{OW14a} working with integer LRS, and is not needed for the proof.}.
\end{theorem}
\begin{proof}
  In~\cite{OW14a}, it is shown that deciding Ultimate Positivity of
  the homogeneous LRS of order \(6\) given by
	\begin{equation*}
	  u_n=r\sin n\theta-n(1-\cos n\theta)\text{ and }v_n=-r\sin n\theta-n(1-\cos n\theta)
	\end{equation*}
  for every \(r\in \QQ\) such that \(r>0\) and \(\theta\in (0,2\pi)\)
  such that \(e^{i\theta}\) has rational real and imaginary parts
  would allow one to compute \(L_\infty(\theta/2\pi)\).
	
  We observe that both sequences \(u_n\) and \(v_n\) fall under the
  premise of Proposition~\ref{prop:inhomo}. Thus, by applying
  Proposition~\ref{prop:inhomo}, we obtain an equivalent
  inhomogeneous LRS of order \(5\), concluding the proof.
\end{proof}

A similar proof, using the results of~\cite{OW14a}, gives us also the
following theorem.
\begin{theorem}
  If Positivity is decidable for inhomogeneous rational LRS of order
  at least 5 then there is an algorithm that computes the
  approximation type of any number \(\theta/2\pi\) such that
  \(e^{i\theta}\) has rational real and imaginary parts.
\end{theorem}

\bibliography{divergence_main}

\begin{thebibliography}{10}

\bibitem{AAOW15}
S.~Akshay, Timos Antonopoulos, Jo{\"{e}}l Ouaknine, and James Worrell.
\newblock Reachability problems for {M}arkov chains.
\newblock {\em Inf. Process. Lett.}, 115(2):155--158, 2015.

\bibitem{Allender}
Eric Allender, Peter B{\"{u}}rgisser, Johan Kjeldgaard{-}Pedersen, and
  Peter~Bro Miltersen.
\newblock On the complexity of numerical analysis.
\newblock {\em {SIAM} J. Comput.}, 38(5):1987--2006, 2009.

\bibitem{baker1993logarithmic}
Alan Baker and Gisbert W{\"u}stholz.
\newblock Logarithmic forms and group varieties.
\newblock {\em J. reine angew. Math}, 442(19-62):3, 1993.

\bibitem{Bau70}
W.~J. Baumol.
\newblock {\em Economic Dynamics. An Introduction}.
\newblock Macmillan, 1970.

\bibitem{CRB13}
Jacek Bochnak, Michel Coste, and Marie-Fran{\c{c}}oise Roy.
\newblock {\em Real algebraic geometry}, volume~36.
\newblock Springer Science \& Business Media, 2013.

\bibitem{bra06}
Mark Braverman.
\newblock Termination of integer linear programs.
\newblock In {\em Computer Aided Verification, 18th International Conference,
  {CAV} 2006, Seattle, WA, USA, August 17-20, 2006, Proceedings}, pages
  372--385, 2006.

\bibitem{cassels1965introduction}
John~W.S. Cassels.
\newblock {\em An Introduction to Diophantine Approximation}.
\newblock Cambridge University Press, 1965.

\bibitem{cohen2013course}
Henri Cohen.
\newblock {\em A course in computational algebraic number theory}, volume 138.
\newblock Springer Science \& Business Media, 2013.

\bibitem{cusick1989markoff}
Thomas~W Cusick and Mary~E Flahive.
\newblock {\em The Markoff and Lagrange spectra}.
\newblock Number~30. American Mathematical Soc., 1989.

\bibitem{EPIW03}
Graham Everest, Alfred~J. van~der Poorten, Igor~E. Shparlinski, and Thomas
  Ward.
\newblock {\em Recurrence Sequences}, volume 104 of {\em Mathematical surveys
  and monographs}.
\newblock American Mathematical Society, 2003.

\bibitem{Eve84}
J.-H. Evertse.
\newblock On sums of {S}-units and linear recurrences.
\newblock {\em Compositio Math.}, 53(2):225--244, 1984.

\bibitem{masser1988linear}
David~W Masser.
\newblock Linear relations on algebraic groups.
\newblock {\em New Advances in Transcendence Theory}, pages 248--262, 1988.

\bibitem{Mig75}
M.~Mignotte.
\newblock A note on linear recursive sequences.
\newblock {\em J. Austral. Math. Soc.}, 20(2):242--244, 1975.

\bibitem{niven2008diophantine}
Ivan~Morton Niven.
\newblock {\em Diophantine approximations}.
\newblock Courier Corporation, 2008.

\bibitem{OWeffective}
Jo{\"{e}}l Ouaknine and James Worrell.
\newblock Effective positivity problems for simple linear recurrence sequences.
\newblock {\em CoRR}, abs/1309.1550, 2013.
\newblock URL: \url{http://arxiv.org/abs/1309.1550}.

\bibitem{OW14b}
Jo{\"{e}}l Ouaknine and James Worrell.
\newblock On the positivity problem for simple linear recurrence sequences,.
\newblock In {\em Automata, Languages, and Programming - 41st International
  Colloquium, {ICALP} 2014, Copenhagen, Denmark, July 8-11, 2014, Proceedings,
  Part {II}}, pages 318--329, 2014.

\bibitem{OW14a}
Jo{\"{e}}l Ouaknine and James Worrell.
\newblock Positivity problems for low-order linear recurrence sequences.
\newblock In {\em Proceedings of the Twenty-Fifth Annual {ACM-SIAM} Symposium
  on Discrete Algorithms, {SODA} 2014, Portland, Oregon, USA, January 5-7,
  2014}, pages 366--379, 2014.

\bibitem{OW14c}
Jo{\"{e}}l Ouaknine and James Worrell.
\newblock Ultimate positivity is decidable for simple linear recurrence
  sequences.
\newblock In {\em Automata, Languages, and Programming - 41st International
  Colloquium, {ICALP} 2014, Copenhagen, Denmark, July 8-11, 2014, Proceedings,
  Part {II}}, pages 330--341, 2014.

\bibitem{OW15}
Jo{\"{e}}l Ouaknine and James Worrell.
\newblock On linear recurrence sequences and loop termination.
\newblock {\em {SIGLOG} News}, 2(2):4--13, 2015.

\bibitem{renegar1992computational}
James Renegar.
\newblock On the computational complexity and geometry of the first-order
  theory of the reals. part i: Introduction. preliminaries. the geometry of
  semi-algebraic sets. the decision problem for the existential theory of the
  reals.
\newblock {\em Journal of symbolic computation}, 13(3):255--299, 1992.

\bibitem{SS83}
T.~N. Shorey and C.~L. Stewart.
\newblock On the {D}iophantine equation $ax^{2t} + bx^ty + cy^2 = d$ and pure
  powers in recurrence sequences.
\newblock {\em Math. Scand.}, 52(1):24--36, 1983.

\bibitem{tarski1951decision}
Alfred Tarski.
\newblock A decision method for elementary algebra and geometry.
\newblock {\em Bulletin of the American Mathematical Society}, 59, 1951.

\bibitem{PS91}
A.~J. van~der Poorten and H.~P. Schlickewei.
\newblock {\em J. Austral. Math. Soc. Ser. A}, 51(1):154--170, 1991.

\end{thebibliography}

\appendix
\section{Complete Proofs of Section~\ref{sec: divergence}}
\label{app: proofs divergence}
Before proceeding with the complete proofs of
Theorems~\ref{thm:divergence 5} and \ref{thm:divergence simple 8}, we
provide some additional mathematical tools,
\subsection{Zero Dimensionality Results}
The following zero-dimensionality lemmas are proved
in~\cite{OWeffective}.
\begin{lemma}
	\label{lem: zero dim}
	Let \(a_1, \ldots, a_m \in \RR\) and \(\phi_1, \ldots, \phi_m \in
        \RR\) be two collections of \(m\) real numbers, for \(m \ge 1\),
        with each of the \(a_i\) non-zero, and let
        \(l_1, \ldots l_m \in \ZZ\) be integers.  Define
        \(f, g: \RR^m \rightarrow \RR\) by setting
        \(f(x_1, \ldots, x_m) = \sum_{i=1}^m a_i cos(x_i + \phi_i)\) and
        \(g(x_1, \ldots, x_m) = \sum_{i=1}^m l_i x_i\).  Assume that
        \(g(x_1, \ldots, x_m)\) is not of the form \(l(x_i \pm x_j)\) for
        some non-zero \(l \in \ZZ\) and indices \(i \ne j\).  Let
        \(\psi \in \RR\).
	
	Then the function \(f\), subject to the constraint
        \(g(x_1, \ldots, x_m) = \psi\), achieves its minimum only
        finitely many times over the domain \(\left[0, 2\pi \right)^m\).
	
\end{lemma}

\begin{lemma}
  \label{lem: zero dim LRS}
  Let \(\tup{u_n}\) be a non-degenerate simple LRS with dominant
  characteristic roots \(\rho \in \RR\) and
  \(\gamma_1, \bar{\gamma}_1, \ldots, \gamma_m, \bar{\gamma}_m \in \CC
  \setminus \RR\).  Write \(\lambda_i = \gamma_i / \rho\) for
  \(1 \le i \le m\), and let
  \(L = \{ (v_1, \ldots, v_m) \in \ZZ^m : \lambda_1^{v_1} \cdots
  \lambda_m^{v_m} = 1 \}\).  Let
  \(\{\Bell_1, \ldots , \Bell_{m-1} \}\) be a basis for \(L\) of
  cardinality \(m-1\), and write
  \(\Bell_j = (\ell_{j,1}, \ldots, \ell_{j,m})\) for
  \(1 \le j \le m-1\). Let

  \begin{equation}
    M =\left( \begin{array}{ccccc}
        l_{1,1} & l_{1,2} & \cdots & l_{1,m-1} & l_{1,m}\\
        l_{2,1} & l_{2,2} & \cdots & l_{2,m-1} & l_{2,m}\\
        \vdots & \vdots & \ddots & \vdots & \vdots\\
        l_{m-1,1} & l_{m-1,2} & \cdots & l_{m-1,m-1} & l_{m-1,m}
        \end{array} \right)
  \end{equation}
	
  Let \(a_1, \ldots, a_m \in \RR\) and \(\phi_1, \ldots, \phi_m\) be
  two collections of \(m\) real numbers, with each of the \(a_i\)
  non-zero, and let
  \(\boldsymbol q = (q_1, \ldots, q_{m-1}) \in \ZZ^{m-1}\) be a column
  vector of \(m-1\) integers.  Let us further write
  \(\boldsymbol x = (x_1, \ldots, x_m)\) to denote a column vector of
  \(m\) real-valued variables.
	
  Then the function
  \(f(x_1, \ldots, x_m) = \sum_{i=1}^{m} a_i cos(x_i + \phi_i)\),
  subject to the constraint \(M \boldsymbol x = 2 \pi \boldsymbol q\),
  achieves its minimum at only finitely many points over the domain
  \(\left[ 0, 2\pi \right)^m\).
\end{lemma}

\subsection{Proof of Theorem~\ref{thm:divergence 5}}
We initially prove Theorem~\ref{thm:divergence 5} for homogeneous
LRS. We then show how to handle the inhomogeneous case, using
Property~\ref{prop:homogenization eigenvalues}.

Consider an LRS \(\mathbf{u}=\tup{u_n}_{n=1}^{\infty}\) of order
\(k \le 5\) with dominant modulus \(\rho\), and let \(\epsilon>0\) be a
rational number.  First, we note that without loss of generality, we
can assume \(\mathbf{u}\) is non-degenerate, as we may decompose a
degenerate sequence and recast analysis at lower orders.  We also note
that if \(\rho<1\), then \(|u_n|\to 0\) as \(n\to \infty\), and in
particular the sequence does not diverge. Thus, we may assume
\(\rho\ge 1\).

By Lemma~\ref{lem:braverman}, if \(\mathbf{u}\) does not have a real
positive dominant root, then \(u_n \not \rightarrow \infty\). Thus, we
may assume a real positive dominant root.  Note that all other
dominant roots must be complex, and come in conjugate pairs, since if
\(-\rho\) were a root, then \(\mathbf{u}\) would be degenerate.

Writing \(u_n\) as an exponential polynomial and dividing by \(\rho^n\),
we have
\begin{eqnarray}
\label{appeq: normalized exp poly}
  \frac{u_n}{\rho^n} = A(n) + \sum_{i=1}^m \left( C_i(n)\lambda_i^n + \conj{C_i(n)}\conj{\lambda}_i^n \right) + r(n),
\end{eqnarray}
where \(A\) is a real polynomial, \(C_i\) are non-zero complex
polynomials, \(\rho\lambda_i\) and \(\rho\conj{\lambda}_i\) are conjugate
pairs of non-real dominant roots of \(\mathbf{u}\), and \(r\) is an
exponentially decaying function (possibly identically zero).  We can
assume that either \(A\not\equiv 0\) or \(m\neq 0\). Indeed, otherwise we
can consider the LRS \(\tup{\rho^n r(n)}_{n=1}^\infty\), which is of
lower order than \(\mathbf{u}\).

We proceed to decide divergence by a case analysis of
Equation~\eqref{appeq: normalized exp poly}.
\paragraph*{Case 1: \(\rho=1\).}
Note that in this case, \(\frac{u_n}{\rho^n}=u_n\).  If \(A(n)\) is a
constant, then it does not affect the divergence of \(\mathbf{u}\). We
claim that \(u_n\not\to \infty\). Indeed, by Lemma~\ref{lem:braverman},
the expression
\(\sum_{i=1}^m \left( C_i(n) \lambda_i^n +
  \conj{C_i(n)}\conj{\lambda}_i^n \right)\) becomes negative infinitely
often (regardless of whether \(C_i\) are constants or polynomials),
whereas the effect of \(r(n)\) is exponentially decreasing. Thus,
\(\mathbf{u}\) does not diverge.

If \(A(n)\) is not a constant, then \(m\le 1\). If \(m=0\), then clearly
\(u_n\to \infty\) iff the leading coefficient of \(A(n)\) is positive.
Otherwise, if \(m=1\), then \(C_1\) is a constant, and thus
\(|C_1\lambda_1^n+\conj{C_1}\conj{\lambda_1}^n|\le 2|C_1|\), and again
\(u_n\to \infty\) iff the leading coefficient of \(A(n)\) is positive.

Recall that since \(\rho=1\), then if \(\mathbf{u}\) diverges, there exist
\(N,k\in \NN\) such that \(u_n\ge n^k\) for all \(n>N\). We now show how to
effectively compute \(N\) and \(k\).

From Proposition~\ref{prop: growth r(n)}, we can compute in polynomial
time \(\epsilon\in (0,1)\) and \(N_1\in \NN\) such that
\(r(n)<(1-\epsilon)^n<1\) for all \(n>N_1\). We thus have that
\(u_n\ge A(n)-|C_1|-1\), and we can easily compute \(N_2\in \NN\) and
\(a\in \QQ\) (depending on the coefficients of \(A(n)\)) such that for all
\(n>N_2\) we have \(A(n)-|C_1|-1\ge an^k\), where \(k\) is the degree of
\(A(n)\), namely 1 or 2. Taking \(N=\max\set{N_1,N_2}\), we conclude this
case.

\paragraph*{Case 2: \(\rho>1\) and there exists a non-constant \(C_i\).}
In this case, \(m=1\), \(C_1\) is linear, and \(A(n)\) is constant.  Let
\(C_1\) have leading coefficient \(b \ne 0\). By
Lemma~\ref{lem:braverman}, there exists \(\epsilon > 0\) such that
\(b\lambda^n + \conj{b}\conj{\lambda}^n < -\epsilon\) infinitely often.
Then \(C_1(n) \lambda_1^n + \conj{C_1(n)} \conj{\lambda}_1^n\) (and
hence \(u_n\)) is unbounded below, so \(u_n \not \to \infty\).

\paragraph*{Case 3: \(\rho>1\) and every \(C_i\) is a nonzero constant.}
In this case, \(m \le 2\). In the following, we set \(m=2\), as the cases
where \(m<2\) are similar and slightly simpler.\footnote{One may notice
  that taking \(m=2\) means that some of the cases we handle actually
  require order \(6\), e.g., when \(A(n)\) is linear and \(m=2\). Still, the
  analysis covers all possible cases of order \(5\).}

Let \(L=\set{(v_1,v_2)\in \ZZ^2: \lambda_1^{v_1}\lambda_2^{v_2}=1}\), and
let \(\set{\Bell_1, \ldots , \Bell_p }\) be a basis
for \(L\) of cardinality \(p\). Write
\(\Bell_{q}=(\ell_{q,1},\ell_{q,2})\) for \(1\le q\le p\). From
Theorem~\ref{thm: basis mult relations}, such a basis can be computed
in polynomial time, and moreover, each \(\ell_{q,j}\) may be assumed
to have magnitude polynomial in \(\norm{\mathbf{u}}\).

Consider the set
\(\TT=\{(z_1,z_2)\in \CC^2: |z_1|=|z_2|=1 \text{ and for each } 1\le
q\le p,\) \( z_1^{\ell_{q,1}}z_2^{\ell_{q,2}}=1\}\).

Define \(h:\TT\to \RR\) by setting
\(h(z_1,z_2)=\sum_{i=1}^2 (C_iz_i+\conj{C_i}\conj{z_i})\), so that for
every \(n\in \NN\),
\(\frac{u_n}{\rho^n}=A(n)+h(\lambda_1^n,\lambda_2^n)+r(n)\). Recall that
the set \(\set{(\lambda_1^n,\lambda_2^n):n\in \NN}\) is a dense subset
of \(\TT\). Since \(h\) is continuous, it follows that
\(\inf\set{h(\lambda_1^n,\lambda_2^n): n\in \NN}=\min h|_\TT=\mu\) for
some \(\mu\in \RR\).

We now claim that \(\mu\) is an algebraic number, computable in
polynomial time, with \(\norm{\mu}=\norm{\mathbf{u}}^{O(1)}\).  We can
represent \(\mu\) via the following formula \(\tau(y)\):
\begin{equation*}
  \exists (\zeta_1,\zeta_2)\in \TT:\ [h(\zeta_1,\zeta_2)=y \wedge \forall (z_1,z_2)\in \TT, y\le h(z_1,z_2)].
\end{equation*} 
Note that \(\tau(y)\) is not a formula in the first-order theory of the
reals, as it involves complex numbers. However, we can rewrite it as a
sentence in the first-order theory of the reals by representing the
real and imaginary parts of each complex quantity and combining them
using real arithmetic (see~\cite{OW14b} for details). In addition, the
obtained formula \(\tau'(y)\) is of size polynomial in
\(\norm{\mathbf{u}}\). By Theorem~\ref{thm: renegar}, we can then
compute in polynomial time an equivalent quantifier-free formula
\begin{equation*}
  \chi(x)=\bigvee_{i=1}^I\bigwedge_{j=1}^{J_i}h_{i,j}\sim_{i,j}0.
\end{equation*}
Recall that each \(\sim_{i,j}\) is either \(>\) or \(=\). Now \(\chi(x)\) must
have a satisfiable disjunct, and since the satisfying assignment to
\(y\) is unique (namely \(y=\mu\)), this disjunct must comprise at least
one equality predicate. Since Theorem~\ref{thm: renegar} guarantees
that the degree and height of each \(h_{i,j}\) are bounded by
\(\norm{\mathbf{u}}^{O(1)}\) and \(2^{\norm{\mathbf{u}}^{O(1)}}\)
respectively, we immediately conclude that \(\mu\) is an algebraic
number and with \(\norm{\mu}=\norm{\mathbf{u}}^{O(1)}\).

We now split the analysis into several cases.

\(\bullet\) If \(A(n)\) is linear with negative leading coefficient, or if
\(A\) is a constant and \(A+\mu<0\), then \(u_n\) is unbounded from below,
and in particular \(u_n \not \rightarrow \infty\).

\(\bullet\) If \(A(n)\) is linear with positive leading coefficient, or if
\(A\) is a constant and \(A+\mu>0\), we can compute in polynomial time
\(N_0 \in \NN\) and a rational \(\epsilon_0>0\) such that
\(A(n) + \mu>2\epsilon_0\) for all \(n>N_0\). By Proposition~\ref{prop:
  growth r(n)}, we can also compute in polynomial time \(N_1\in \NN\)
and \(\epsilon_1\in (0,1)\) such that \(|r(n)|<(1-\epsilon_1)^n\) for all
\(n>N_1\). Taking \(N_2\ge \log_{1-\epsilon_1}\epsilon_0\), we have that
for all \(n>\max\set{N_0,N_1,N_2}\), \(|r(n)|< \epsilon_0\), and thus
\begin{equation*}
  \frac{u_n}{\rho^n} =  A(n) + h(\lambda_1, \ldots, \lambda_m)+r(n)
  \ge A(n) + \mu-\epsilon_0
  > 2\epsilon_0-\epsilon_0=\epsilon_0\, .
\end{equation*}
Thus we have \(u_n \ge \epsilon_0 \rho^n\) for all \(n>\max
\set{N_0,N_1,N_2}\) and hence we have effective growth bounds in this case.

%%%%%%%
\(\bullet\) If \(A\) is a constant and \(A+\mu=0\), things are more involved. Let \(\lambda_j=e^{i\theta_j}\) and \(C_j=|C_j|e^{i\phi_j}\) for \(1\le j\le 2\). From Equation~\eqref{appeq: normalized exp poly} we have 
\begin{equation*}
  \frac{u_n}{\rho^n}=A+\sum_{j=1}^2 2|C_j|\cos(n\theta_j+\phi_j)+r(n).
\end{equation*}
We further assume that all the \(C_j\) are non-zero. Indeed, if this
does not hold, we can recast our analysis in lower dimension.

We now claim that \(h\) achieves its minimum \(\mu\) only finitely many
times over \(\TT\). To establish this claim, we proceed according to the
cardinality \(p\) of the basis
\(\set{\Bell_1, \ldots , \Bell_p }\) of \(L\):

(i) We first consider the case in which \(p=1\), and handle the case
\(p=0\) immediately afterwards. Let
\(\Bell_1=(\ell_{1,1},\ell_{1,2})\in \ZZ^2\) be the sole vector
spanning \(L\). For \(x\in \RR\), recall that we denote by \([x]_{2\pi}\)
the distance from \(x\) to the closest integer multiple of \(2\pi\).

Write
\begin{equation*}
  R=\set{(x_1,x_2)\in [0,2\pi)^2: [\ell_{1,1}x_1+\ell_{1,2}x_2]_{2\pi}=0}.
\end{equation*}
Clearly, for any \((x_1,x_2)\in [0,2\pi)^2\), we have \((x_1,x_2)\in R\)
iff \((e^{ix_1},e^{i x_2})\in \TT\).  Define \(f:\RR^2\to \RR\) by setting
\begin{equation*}
  f(x_1,x_2)=\sum_{j=1}^2 2|C_j|\cos(x_j+\phi_j).
\end{equation*}
Clearly, for all \((x_1,x_2)\in [0,2\pi)^2\) we have
\(f(x_1,x_2)=h(e^{ix_1},e^{ix_2})\), and therefore the minimal values of
\(f\) over \(\RR\) are in one-to-one correspondence with those of \(h\)
over \(\TT\).

Define \(g:\RR^2\to \RR\) by setting
\begin{equation*}
  g(x_1,x_2)=\ell_{1,1}x_1+\ell_{1,2}x_2.
\end{equation*}
Note that \(g(x_1,x_2)\) cannot be of the form \(\ell(x_i-x_j)\), for
nonzero \(\ell\in \ZZ\) and \(i\neq j\), otherwise
\(\lambda_i^\ell\lambda_j^{-\ell}=1\), i.e., \(\lambda_i/\lambda_j\) would
be a root of unity, contradicting the non-degeneracy of
\(\mathbf{u}\). Likewise, \(g\) cannot be of the form \(\ell(x_i+x_j)\),
otherwise \(\lambda_i/\conj{\lambda}_j\) would be a root of unity.

Finally, observe that for \((x_1,x_2)\in [0,2\pi)^2\), we have
\((x_1,x_2)\in R\) iff \(\ell_{1,1}x_1+\ell_{1,2}x_2=2\pi q\) for some
\(q\in \ZZ\) with \(|q|\le |\ell_{1,1}|+|\ell_{1,2}|\). For each of these
finitely many \(q\), we can invoke Lemma~\ref{lem: zero dim} with \(f,g,\)
and \(\psi=2\pi q\), to conclude that \(f\) achieves its minimum \(\mu\)
finitely many times over \(R\), and therefore that \(h\) achieves the same
minimum finitely many times over \(\TT\).

The case \(p=0\), i.e., in which there are no non-trivial integer
multiplicative relationships among \(\lambda_1,\lambda_2\), is now a
special case of the above, where we have \(\ell_{1,1}=\ell_{1,2}\).

(ii) We observe that the case \(p=2\) cannot occur: indeed, a basis for
\(L\) of dimension \(2\) would immediately entail that every \(\lambda_j\)
is a root of unity.

This concludes the proof of the claim that \(h\) achieves its minimum at
a finite number of points
\(Z=\set{(\zeta_1,\zeta_2)\in \TT: h(\zeta_1,\zeta_2)=\mu}\).

We concentrate on the set \(Z_1\) of first coordinates of \(Z\). Write
\begin{align*}
\tau_1(x)=\exists z_1 (\re(z_1)=x\wedge z_1\in Z_1),\\
\tau_2(y)=\exists z_1 (\im(z_1)=y\wedge z_1\in Z_1).
\end{align*}
Similarly to our earlier construction, \(\tau_1(x)\) is equivalent to a
formula \(t'_1(x)\) in the in the first-order theory of the reals, over
a bounded number of real variables, with
\(\norm{\tau'_1(x)}=\norm{\mathbf{u}}^{O(1)}\).  Thanks to
Theorem~\ref{thm: renegar}, we then obtain an equivalent
quantifier-free formula
\begin{equation*}
  \chi_1(x)=\bigvee_{i=1}^I\bigwedge_{j=1}^{J_i}h_{i,j}\sim_{i,j}0.
\end{equation*}
Note that since there can only be finitely many \(\hat{x}\in \RR\) such
that \(\chi_1(\hat{x})\) holds, each disjunct of \(\chi_1(\hat{x})\) must
comprise at least one equality predicate, or can otherwise be entirely
discarded as having no solution.  A similar exercise can be carried
out with \(\tau_2(y)\) to obtain \(\chi_2(y)\). The bounds on the degree
and height of each \(h_{i,j}\) in \(\chi_1(x)\) and \(\chi_2(y)\) then
enables us to conclude that any \(\zeta_1=\hat{x}+i \hat{y}\in Z_1\) is
algebraic, and moreover satisfies
\(\norm{\zeta_1}= \norm{\mathbf{u}}^{O(1)}\).  In addition, bounds on
\(I\) and \(J_i\) guarantee that the cardinality of \(Z_1\) is at most
polynomial in \(\norm{\mathbf{u}}\).

Since \(\lambda_1\) is not a root of unity, for each \(\zeta_1\in Z_1\)
there is at most one value of \(n\) such that \(\lambda_1^n=\zeta_1\).
Theorem~\ref{thm: basis mult relations} then entails that this value
(if it exists) is at most \(M = \norm{\mathbf{u}}^{O(1)}\), which we can
take to be uniform across all \(\zeta_1\in Z_1\).  We can now invoke
Lemma~\ref{lem: simple baker} to conclude that, for \(n > M\), and
for all \(\zeta_1\in Z_1\), we have
\begin{equation}
\label{appeq: baker usage zeta}
|\lambda_1^n-\zeta_1|>\frac{1}{n^{\norm{\mathbf{u}}^{D}}},
\end{equation}
where \(D\in \NN\) is some absolute constant.

Let \(b>0\) be minimal such that the set
\begin{equation*}
  \set{z_1\in \CC: |z_1|=1 \text{ and, for all } \zeta_1\in Z_1, |z_1-\zeta_1|\ge \frac1b}
\end{equation*} 
is non empty. Thanks to our bounds on the cardinality of \(Z_1\), we can
use the first-order theory of the reals, together with
Theorem~\ref{thm: renegar}, to conclude that \(b\) is algebraic and
\(\norm{b}=\norm{\mathbf{u}}^{O(1)}\).

Define the function \(g:[b,\infty)\to \RR\) as follows:
\begin{align*}
  g(x)=\min\{h(z_1,z_2)-\mu: (z_1,z_2)\in \TT \text{ and for all } \zeta_1\in Z_1, |z_1-\zeta_1|\ge \frac{1}{x}\}.
\end{align*}
It is clear that \(g\) is continuous and \(g(x) > 0\) for all
\(x\in [b,\infty)\).  Moreover, \(g\) can be translated in polynomial time
into a function in the first-order theory of the reals over a bounded
number of variables.  It follows from Proposition 2.6.2
of~\cite{CRB13} (invoked with the function \(1/g\)) that there is a
polynomial \(P\in \ZZ[x] \) such that, for all \(x\in [b,\infty)\),
\begin{equation}
\label{appeq: bound on g}
g(x)\ge \frac{1}{P(x)}.
\end{equation}
Moreover, and examination of the proof of \cite[Prop. 2.6.2]{CRB13}
reveals that \(P\) is obtained through a process which hinges on
quantifier elimination.  By Theorem~\ref{thm: renegar}, we are
therefore able to conclude that \(\norm{P}=\norm{\mathbf{u}}^{O(1)}\), a
fact which relies among others on our upper bounds for \(\norm{b}\).

By Proposition~\ref{prop: growth r(n)} we can find \(\epsilon\in (0,1)\)
and \(N=2^{\norm{\mathbf{u}}^{O(1)}}\) such that for all \(n>N\), we have
\(|r(n)|<(1-\epsilon)^n\), and moreover
\(1/\epsilon=2^{\norm{\mathbf{u}}^{O(1)}}\).  In addition, by
Proposition~\ref{prop: bound inverse poly}, there is
\(N'=2^{\norm{\mathbf{u}}^{O(1)}}\) such that for every \(n\ge N'\)
\begin{equation}
\label{appeq: bound on P}
\frac{1}{2P(n^{\norm{\mathbf{u}}^D})}>(1-\epsilon)^n.
\end{equation}
Combining Equations~\eqref{appeq: normalized exp poly}--\eqref{appeq:
  bound on P}, we get
\begin{align*}
  \frac{u_n}{\rho^n}&=A+h(\lambda_1^n,\lambda_2^n)+r(n)\\
                    &\ge-\mu+h(\lambda_1^n,\lambda_2^n)-(1-\epsilon)^n\\
                    &\ge g(n^{\norm{\mathbf{u}}^D})-(1-\epsilon)^n\\
                    &\ge \frac{1}{P(n^{\norm{\mathbf{u}}^D})}-(1-\epsilon)^n\\
                    &= \frac{1}{2P(n^{\norm{\mathbf{u}}^D})}+\frac{1}{2P(n^{\norm{\mathbf{u}}^D})}-(1-\epsilon)^n\\
                    &\ge \frac{1}{2P(n^{\norm{\mathbf{u}}^D})}
\end{align*}
provided \(n>\max\set{M,N,N'}\). We thus have that \(\frac{u_n}{\rho^n}\)
is eventually lower bounded by an inverse polynomial and hence we have
effective growth bounds in this case.

This concludes the decidability of divergence and computability of
effective bounds on divergence for homogeneous LRS of order at most
\(5\).

It remains to show how to handle inhomogeneous LRS of order at
most\footnote{In fact, by property~\ref{prop:homogenization
    eigenvalues}, LRS of order at most \(4\) can be handled by
  homogenization. Thus, it is enough to handle exactly order 5.} \(5\).
Consider an inhomogeneous LRS \(\mathbf{v}=\tup{v_n}_{n=1}^\infty\) of
order 5, and let \(\mathbf{u}=\homLRS(\mathbf{v})\). Consider the
dominant modulus \(\rho\) of \(u_n\). If \(\rho>1\), then by
property~\ref{prop:homogenization eigenvalues} the exponential
polynomial of \(\frac{u_n}{\rho^n}\) is the same as that in
Equation~\eqref{appeq: normalized exp poly}. Thus, we can proceed with
the case analysis of Case 2 and Case 3 without change.  If \(\rho=1\),
things become more involved. Consider the exponential polynomial
\begin{eqnarray}
\label{appeq: exp poly homogenized}
  u_n = A(n) + \sum_{i=1}^m \left( C_i(n)(\lambda_i^n) + \conj{C_i(n)}(\conj{\lambda}_i^n) \right) + r(n)
\end{eqnarray}
where \(|r(n)|\) is exponentially decaying and the \(\lambda_i\) are
characteristic roots of modulus \(1\).

If \(A(n)\) is constant, or if \(A(n)\) is not a constant and all the
\(C_i\) are constants (if there are any), then the same analysis of Case
1 applies here, \emph{mutatis-mutandis}. Otherwise, the only possible
case is where \(A(n)\) is linear, \(m=1\), \(C_1(n)\) is linear, and
\(r(n)\equiv 0\). Indeed, this corresponds to the case where the
characteristic roots of \(u_n\) are \(1,\lambda,\conj{\lambda}\), each
with multiplicity \(2\). Let \(A(n)=a_1 n+ b_1\) and \(C_1(n)=a_2 n+b_2\),
then we can write
\begin{equation*}
  u_n=a_1 n+ b_1+ (a_2 n+ b_2)\lambda^n+ (\conj{a_2} n +
  \conj{b_2})\conj{\lambda}^n=n(a_1+a_2\lambda^n +
  \conj{a_2}\conj{\lambda^n})+(b_1+b_2\lambda^n +
  \conj{b_2}\conj{\lambda^n}).
\end{equation*}
Since \(|(b_1+b_2\lambda^n+\conj{b_2}\conj{\lambda^n})|\) is bounded,
then \(u_n\) diverges iff
\(n(a_1+a_2\lambda^n+\conj{a_2}\conj{\lambda^n})\) diverges. Let
\(\theta=\arg \lambda\) and \(\phi=\arg a_2\). We have
\(n(a_1+a_2\lambda^n+\conj{a_2}\conj{\lambda^n})=n(a_1+2|a_2|\cos(n\theta+\phi))\).

Again, we split into cases.

\(\bullet\) If \(a_1>2|a_2|\), we have that \(u_n\) diverges. Then we can
compute in polynomial time a rational \(\epsilon>0\) and \(N\in \NN\) such
that \(a_1-2|a_2|>\epsilon\) and \(n(a_1+2|a_2|)-(b_1-2|b_2|)>\epsilon n\)
for all \(n>N\). We then have that \(u_n>\epsilon n\) for all \(n>N\), thus
concluding effective decidability of divergence in this case.

\(\bullet\) If \(a_1<2|a_2|\), then \(u_n\) does not diverge, as it becomes
negative infinitely often.

\(\bullet\) The remaining case is when \(a_1=2|a_2|\), where the
expression above becomes \(na_1(1+\cos(n\theta+\phi))\). We show that in
this case, \(u_n\) does not diverge.  By Taylor approximation, for every
\(x\in (-\pi,\pi]\) it holds that \(1-\cos(x)\le \frac{x^2}{2}\).  For
\(n\in \NN\), write \(\Lambda(n)=n\theta+\phi-(2j+1)\pi\), where
\(j\in \ZZ\) is the unique integer such that \(-\pi<\Lambda(n)\le \pi\).
We now have that
\begin{align*}
  na_1(1+\cos(n\theta+\phi)) = na_1(1-\cos(n\theta+\phi+\pi))= na_1(1-\cos(\Lambda(n)))<na_1\frac{\Lambda(n)^2}{2}.
\end{align*}
By Dirichlet's Approximation Theorem, we have that
\(|\Lambda(n)|<\frac{t}{n}\) for infinitely many values of \(n\), where
\(t\) is a constant depending on \(\phi\). Thus, we have
\(na_1\frac{\Lambda(n)^2}{2}<\frac{a_1t^2}{2n}\).  It follows that \(u_n\)
is infinitely often bounded by a constant, and does not diverge.

\subsection{Proof of Theorem~\ref{thm:divergence simple 8}}
As in the proof of Theorem~\ref{thm:divergence 5}, we start by
considering the homogeneous case, and we let \(\tup{u_n}\) be a
non-degenerate simple LRS of order \(d \le 8\) with a real positive
dominant characteristic root \(\rho\ge 1\).

As before, we write 
\begin{equation}
\label{appeq: simple exp poly}
\frac{u_n}{\rho^n}=a+\sum_{i=1}^m (c_i\lambda_i^n+\conj{c_i}\conj{\lambda_i^n})+r(n)
\end{equation}
with \(a\in \RR\), \(c_i\in \CC\setminus \RR\) for every \(1\le i\le m\),
and \(|r(n)|\) exponentially decaying. Note that since \(d\le 8\) and
\(a\in \RR\), it follows that \(0\le m\le 3\).  In the following, we
consider the case where \(m=3\). The cases where \(m<3\) are very similar
and slightly simpler, and are therefore omitted.

Observe that if \(\rho=1\), the sequence \(u_n\) is bounded, and therefore
does not diverge. We hence assume \(\rho>1\).

Let
\(L=\set{(v_1,\ldots,v_m)\in \ZZ^m: \lambda_1^{v_1}\cdots
  \lambda_m^{v_m}=1}\), and let
\(\set{\Bell_1, \ldots , \Bell_p }\) be a basis for
\(L\) of cardinality \(p\). Write
\(\Bell_{q}=(\ell_{q,1},\ldots,\ell_{q,m})\) for \(1\le q\le
p\). From Theorem~\ref{thm: basis mult relations}, such a basis can be
computed in polynomial time, and moreover -- each \(\ell_{q,j}\) may be
assumed to have magnitude polynomial in \(\norm{u}\).

Consider the set
\(\TT=\{(z_1,z_2,z_3)\in \CC^3: |z_1|=|z_2|=|z_3|=1 \text{ and for each
} 1\le q\le p,\)
\( z_1^{\ell_{q,1}}z_2^{\ell_{q,2}}z_3^{\ell_{q,3}}=1\}\).

Define \(h:\TT\to \RR\) by setting
\(h(z_1,z_2,z_3)=\sum_{i=1}^3 (c_iz_i+\conj{c_i}\conj{z_i})\), so that
for every \(n\in \NN\),
\(\frac{u_n}{\rho^n}=a+h(\lambda_1^n,\lambda_2^n,\lambda_3^n)+r(n)\). Recall
that the set \(\set{\lambda_1^n,\lambda_2^n,\lambda_3^n:n\in \NN}\) is a
dense subset of \(\TT\). Since \(h\) is continuous, it follows that
\(\inf\set{h(\lambda_1^n,\lambda_2^n,\lambda_3^n): n\in \NN}=\min
h|_\TT=\mu\) for some \(\mu\in \RR\).

We now claim that \(\mu\) is an algebraic number, computable in
polynomial time, with \(\norm{\mu}=\norm{\mathbf{u}}^{O(1)}\).  We can
represent \(\mu\) via the following formula \(\tau(y)\):
\begin{equation*}
\exists (\zeta_1,\zeta_2,\zeta_3)\in \TT : \ [h(\zeta_1,\zeta_2,\zeta_3)
= y \wedge \forall (z_1,z_2,z_3)\in \TT, y\le h(z_1,z_2,z_3)].
\end{equation*} 
Note that \(\tau(y)\) is not a formula in the first-order theory of the
reals, as it involves complex numbers. However, we can rewrite it as a
sentence in the first-order theory of the reals by representing the
real and imaginary parts of each complex quantity and combining them
using real arithmetic (see~\cite{OW14b} for details). In addition, the
obtained formula \(\tau'(y)\) is of size polynomial in
\(\norm{\mathbf{u}}\). By Theorem~\ref{thm: renegar}, we can then
compute in polynomial time an equivalent quantifier-free formula
\begin{equation*}
\chi(x)=\bigvee_{i=1}^I\bigwedge_{j=1}^{J_i}h_{i,j}\sim_{i,j}0.
\end{equation*}
Recall that each \(\sim_{i,j}\) is either \(>\) or \(=\). Now \(\chi(x)\) must
have a satisfiable disjunct, and since the satisfying assignment to
\(y\) is unique (namely \(y=\mu\)), this disjunct must comprise at least
one equality predicate. Since Theorem~\ref{thm: renegar} guarantees
that the degree and height of each \(h_{i,j}\) are bounded by
\(\norm{\mathbf{u}}^{O(1)}\) and \(2^{\norm{\mathbf{u}}^{O(1)}}\)
respectively, we immediately conclude that \(\mu\) is an algebraic
number and with \(\norm{\mu}=\norm{\mathbf{u}}^{O(1)}\).

We now split to cases according to the sign of \(a+\mu\).

\(\bullet\) If \(a+\mu<0\), then \(\mathbf{u}\) is infinitely often
negative, and does not diverge.

\(\bullet\) If \(a+\mu>0\), then \(\mathbf{u}\) diverges, and it remains to
show an effective bound. We can compute in polynomial time a rational
\(\epsilon_0>0\) such that \(a + \mu>2\epsilon_0\). By
Proposition~\ref{prop: growth r(n)}, we can also compute in polynomial
time \(N_1\in \NN\) and \(\epsilon_1\in (0,1)\) such that
\(|r(n)|<(1-\epsilon_1)^n\) for all \(n>N_1\). Taking
\(N_2\ge \log_{1-\epsilon_1}\epsilon_0\), we have that for all
\(n>\max\set{N_1,N_2}\), \(|r(n)|< \epsilon_0\), and
thus
\begin{equation*}
\frac{u_n}{\rho^n} = A(n) + h(\lambda_1, \ldots, \lambda_m)+r(n)
\ge A(n) + \mu-\epsilon_0
> 2\epsilon_0-\epsilon_0=\epsilon_0.
\end{equation*}
Thus \(u_n > \epsilon_0 \rho^n\)
for all \(n>\max\set{N_1,N_2}\) and hence we have effective divergence
bounds in this case.

\(\bullet\) It remains to analyze the case where \(a+\mu=0\). To this end,
let \(\lambda_j=e^{i\theta_j}\) and \(c_j=|c_j|e^{i\phi_j}\) for
\(1\le j\le 3\). From Equation~\eqref{appeq: simple exp poly} we have
\begin{equation*}
  \frac{u_n}{\rho^n}=a+\sum_{j=1}^3 2|c_j|\cos(n\theta_j+\phi_j)+r(n).
\end{equation*}
We further assume that all the \(c_j\) are non-zero. Indeed, if this
does not hold, we can recast our analysis in lower dimension.

We now claim that \(h\) achieves its minimum \(\mu\) only finitely many
times over \(\TT\). To establish this claim, we proceed according to the
cardinality \(p\) of the basis
\(\set{\Bell_1, \ldots , \Bell_p }\) of \(L\):

(i) We first consider the case in which \(p=1\), and handle the case
\(p=0\) immediately afterwards. Let
\(\Bell_1=(\ell_{1,1},\ell_{1,2},\ell_{1,3})\in \ZZ^3\) be the
sole vector spanning \(L\). For \(x\in \RR\), recall that we denote by
\([x]_{2\pi}\) the distance from \(x\) to the closest integer multiple of
\(2\pi\).  Write
\begin{equation*}
  R=\set{(x_1,x_2,x_3)\in [0,2\pi)^3: [\ell_{1,1}x_1+\ell_{1,2}x_2+\ell_{1,3}x_3]_{2\pi}=0}.
\end{equation*}
Clearly, for any \((x_1,x_2,x_3)\in [0,2\pi)^3\), we have
\((x_1,x_2,x_3)\in R\) iff \((e^{ix_1},e^{i x_2},e^{i x_3})\in \TT\).
Define \(f:\RR^3\to \RR\) by setting
\begin{equation*}
  f(x_1,x_2,x_3)=\sum_{j=1}^3 2|c_j|\cos(x_j+\phi_j).
\end{equation*}
Clearly, for all \((x_1,x_2,x_3)\in [0,2\pi)^3\) we have
\(f(x_1,x_2,x_3)=h(e^{ix_1},e^{ix_2},e^{ix_3})\), and therefore the
minimal values of \(f\) over \(\RR\) are in one-to-one correspondence with
those of \(h\) over \(\TT\).

Define \(g:\RR^3\to \RR\) by setting 
\begin{equation*}
  g(x_1,x_2,x_3)=\ell_{1,1}x_1+\ell_{1,2}x_2+\ell_{1,3}x_3.
\end{equation*}
Note that \(g(x_1,x_2,x_3)\) cannot be of the form \(\ell(x_i-x_j)\), for
nonzero \(\ell\in \ZZ\) and \(i\neq j\), otherwise
\(\lambda_i^\ell\lambda_j^{-\ell}=1\), i.e., \(\lambda_i/\lambda_j\) would
be a root of unity, contradicting the non-degeneracy of
\(\mathbf{u}\). Likewise, \(g\) cannot be of the form \(\ell(x_i+x_j)\),
otherwise \(\lambda_i/\conj{\lambda}_j\) would be a root of unity.

Finally, observe that for \((x_1,x_2,x_3)\in [0,2\pi)^3\), we have
\((x_1,x_2,x_3)\in R\) iff
\(\ell_{1,1}x_1+\ell_{1,2}x_2+\ell_{1,3}x_3=2\pi q\) for some \(q\in \ZZ\)
with \(|q|\le |\ell_{1,1}|+|\ell_{1,2}|+|\ell_{1,3}|\). For each of
these finitely many \(q\), we can invoke Lemma~\ref{lem: zero dim} with
\(f,g,\) and \(\psi=2\pi q\), to conclude that \(f\) achieves its minimum
\(\mu\) finitely many times over \(R\), and therefore that \(h\) achieves
the same minimum finitely many times over \(\TT\).

The case \(p=0\), i.e., in which there are no non-trivial integer
multiplicative relationships among \(\lambda_1,\lambda_2,\lambda_3\), is
now a special case of the above, where we have
\(\ell_{1,1}=\ell_{1,2}=\ell_{1,3}=0\).

(ii) We now turn to the case \(p=2\). We have \(\Bell_1=(\ell_{1,1},\ell_{1,2},\ell_{1,3})\in \ZZ^3\) and \(\Bell_2=(\ell_{2,1},\ell_{2,2},\ell_{2,3})\in \ZZ^3\) spanning \(L\). Let \(\boldsymbol x\) denote the column vector \((x_1,x_2,x_3)\), and write
\begin{equation*}
  R=\set{(x_1,x_2,x_3)\in [0,2\pi)^3: [\Bell_1\cdot
  \boldsymbol x]_{2\pi}=0\text{ and }[\Bell_2\cdot
  \boldsymbol x]_{2\pi}=0}.
  \end{equation*}
Define \(f:\RR^3\to \RR\) by setting
\(f(x_1,x_2,x_3)=\sum_{j=1^3}2|c_j|\cos(x_j+\phi_j)\). As before, the
minima of \(f\) over \(R\) are in one-to-one correspondence with those of
\(h\) over \(\TT\).

For \((x_1,x_2,x_3)\in [0,2\pi)^3\), we have
\([\Bell_1\cdot \boldsymbol x]_{2\pi}=0\) and
\([\Bell_2\cdot \boldsymbol x]_{2\pi}=0\) iff there exist
\(q_1,q_2\in \ZZ\), with
\(|q_1|\le |\ell_{1,1}|+|\ell_{1,2}|+|\ell_{1,3}|\) and
\(|q_2|\le |\ell_{2,1}|+|\ell_{2,2}|+|\ell_{2,3}|\) such that
\(\Bell_1\cdot \boldsymbol x=2\pi q_1\) and
\(\Bell_2\cdot \boldsymbol x=2\pi q_2\). For each of theses
finitely many \(\boldsymbol q=(q_1,q_2)\), we can invoke Lemma~\ref{lem:
  zero dim LRS} with \(f\), \(M=\begin{pmatrix}
  \ell_{1,1} & \ell_{1,2} & \ell_{1,3}\\
  \ell_{2,1} & \ell_{2,2} & \ell_{2,3}
\end{pmatrix}\), and \(\boldsymbol q\), to conclude that \(f\) achieves its
minimum \(\mu\) finitely many times over \(R\), and therefore that \(h\)
achieves the same minimum finitely many times over \(\TT\).

(iii) Finally, we observe that the case \(p=3\) cannot occur: indeed, a
basis for \(L\) of dimension \(3\) would immediately entail that every
\(\lambda_j\) is a root of unity.

This concludes the proof of the claim that \(h\) achieves its minimum at
a finite number of points
\(Z=\set{(\zeta_1,\zeta_2,\zeta_3)\in \TT:
  h(\zeta_1,\zeta_2,\zeta_3)=\mu}\).

We concentrate on the set \(Z_1\) of first coordinates of \(Z\). Write
\begin{align*}
\tau_1(x)=\exists z_1 (\re(z_1)=x\wedge z_1\in Z_1),\\
\tau_2(y)=\exists z_1 (\im(z_1)=y\wedge z_1\in Z_1).
\end{align*}
Similarly to our earlier constructions \(\tau_1(x)\) is equivalent to a
formula \(t'_1(x)\) in the in the first-order theory of the reals, over
a bounded number of real variables, with
\(\norm{\tau'_1(x)}=\norm{\mathbf{u}}^{O(1)}\).  Thanks to
Theorem~\ref{thm: renegar}, we then obtain an equivalent
quantifier-free formula
\begin{equation*}
  \chi_1(x)=\bigvee_{i=1}^I\bigwedge_{j=1}^{J_i}h_{i,j}\sim_{i,j}0.
\end{equation*}
Note that since there can only be finitely many \(\hat{x}\in \RR\) such
that \(\chi_1(\hat{x})\) holds, each disjunct of \(\chi_1(\hat{x})\) must
comprise at least one equality predicate, or can otherwise be entirely
discarded as having no solution.  A similar exercise can be carried
out with \(\tau_2(x)\). The bounds on the degree and height of each
\(h_{i,j}\) in \(\chi_1(x)\) and \(\chi_2(y)\) then enable us to conclude
that any \(\zeta_1=\hat{x}+i \hat{y}\in Z_1\) is algebraic, and moreover
satisfies \(\norm{\zeta_1}= \norm{\mathbf{u}}^{O(1)}\).  In addition,
bounds on \(I\) and \(J_i\) guarantee that the cardinality of \(Z_1\) is at
most polynomial in \(\norm{\mathbf{u}}\).

Since \(\lambda_1\) is not a root of unity, for each \(\zeta_1\in Z_1\)
there is at most one value of \(n\) such that \(\lambda_1^n=\zeta_1\).
Theorem~\ref{thm: basis mult relations} then entails that this value
(if it exists) is at most \(M = \norm{\mathbf{u}}^{O(1)}\), which we can
take to be uniform across all \(\zeta_1\in Z_1\).  We can now invoke
Corollary~\ref{lem: simple baker} to conclude that, for \(n > M\), and
for all \(\zeta_1\in Z_1\), we have
\begin{equation}
\label{appeq: baker usage zeta simple}
|\lambda_1^n-\zeta_1|>\frac{1}{n^{\norm{\mathbf{u}}^{D}}},
\end{equation}
where \(D\in \NN\) is some absolute constant.

Let \(b>0\) be minimal such that the set
\begin{equation*}
  \set{z_1\in \CC: |z_1|=1 \text{ and, for all } \zeta_1\in Z_1, |z_1-\zeta_1|\ge \frac1b}
\end{equation*} 
is non empty. Thanks to our bounds on the cardinality of \(Z_1\), we can
use the first-order theory of the reals, together with
Theorem~\ref{thm: renegar}, to conclude that \(b\) is algebraic and
\(\norm{b}=\norm{\mathbf{u}}^{O(1)}\).

Define the function \(g:[b,\infty)\to \RR\) as follows:
\begin{align*}
  g(x)=\min\{h(z_1,z_2,z_3)-\mu: (z_1,z_2,z_3)\in \TT \text{ and for all } \zeta_1\in Z_1, |z_1-\zeta_1|\ge \frac{1}{x}\}.
\end{align*}
It is clear that \(g\) is continuous and \(g(x) > 0\) for all
\(x\in [b,\infty)\).  Moreover, \(g\) can be translated in polynomial time
into a function in the first-order theory of the reals over a bounded
number of variables.  It follows from Proposition 2.6.2
of~\cite{CRB13} (invoked with the function \(1/g\)) that there is a
polynomial \(P\in \ZZ[x] \) such that, for all \(x\in [b,\infty)\),
\begin{equation}
\label{appeq: bound on g simple}
g(x)\ge \frac{1}{P(x)}.
\end{equation}
Moreover, and examination of the proof of \cite[Prop. 2.6.2]{CRB13}
reveals that \(P\) is obtained through a process which hinges on
quantifier elimination.  By Theorem~\ref{thm: renegar}, we are
therefore able to conclude that \(\norm{P}=\norm{\mathbf{u}}^{O(1)}\), a
fact which relies among others on our upper bounds for \(\norm{b}\).

By Proposition~\ref{prop: growth r(n)} we can find \(\epsilon\in (0,1)\)
and \(N=2^{\norm{\mathbf{u}}^{O(1)}}\) such that for all \(n>N\), we have
\(|r(n)|<(1-\epsilon)^n\), and moreover
\(1/\epsilon=2^{\norm{\mathbf{u}}^{O(1)}}\).  In addition, by
Proposition~\ref{prop: bound inverse poly}, there is
\(N'=2^{\norm{\mathbf{u}}^{O(1)}}\) such that for every \(n\ge N'\)
\begin{equation}
\label{appeq: bound on P simple}
\frac{1}{2P(n^{\norm{\mathbf{u}}^D})}>(1-\epsilon)^n.
\end{equation}
Combining Equations~\eqref{appeq: simple exp poly}--\eqref{appeq:
  bound on P simple}, we get
\begin{align*}
  \frac{u_n}{\rho^n}&=a+h(\lambda_1^n,\lambda_2^n,\lambda_3^n)+r(n)\\
                    &\ge-\mu+h(\lambda_1^n,\lambda_2^n,\lambda_3^n)-(1-\epsilon)^n\\
                    &\ge g(n^{\norm{\mathbf{u}}^D})-(1-\epsilon)^n\\
                    &\ge \frac{1}{P(n^{\norm{\mathbf{u}}^D})}-(1-\epsilon)^n\\
                    &= \frac{1}{2P(n^{\norm{\mathbf{u}}^D})}+\frac{1}{2P(n^{\norm{\mathbf{u}}^D})}-(1-\epsilon)^n\\
                    &\ge \frac{1}{2P(n^{\norm{\mathbf{u}}^D})}
\end{align*}
provided \(n>\max\set{M,N,N'}\). We thus have that \(\frac{u_n}{\rho^n}\)
is eventually lower bounded by an inverse polynomial and hence we have
effective divergence bounds in this case.

It remains to show how to handle inhomogeneous LRS of order at most
\(8\).  Consider an inhomogeneous LRS \(\tup{v_n}\) of order at most 8, and
let \(u_n=\homLRS(v_n)\). Observe that by
Property~\ref{prop:homogenization eigenvalues}, \(u_n\) might not be a
simple LRS. However, all its characteristic roots have multiplicity
\(1\), apart from, possibly, the characteristic root 1.

Consider the dominant modulus \(\rho\) of \(u_n\). If \(\rho>1\), then by
property~\ref{prop:homogenization eigenvalues} the exponential
polynomial of \(\frac{u_n}{\rho^n}\) is of the same form as that in
Equation~\eqref{appeq: normalized exp poly}, in the sense that \(r(n)\)
is still exponentially decaying. Thus, we can proceed with the
analysis above without change.  If \(\rho=1\), things become slightly
more involved. Consider the exponential polynomial
\begin{equation}
u_n=A(n)+\sum_{i=1}^m (c_i\lambda_i^n+\conj{c_i}\conj{\lambda_i^n})+r(n)
\end{equation}
where \(A(n)\) is either a constant or a polynomial,
\(c_i\in \CC\setminus \RR\) for every \(1\le i\le m\), \(|r(n)|\) is
exponentially decaying, and \(0\le m\le 4\).  Observe that if \(A(n)\) is
a constant, then \(|u_n|\) is bounded, and so does not diverge. In
particular, if \(m=4\) then it has to be the case that \(A(n)\) is
constant. Thus, it suffices to consider the case where \(m\le 3\) and
\(A(n)\) is a polynomial.

In this case, similarly to Case 1 in the proof of
Theorem~\ref{thm:divergence 5}, we have that \( \mathbf{u}\) diverges
iff the leading coefficient of \(A(n)\) is positive, and in this case
the bounds are effectively computable.

This completes the proof of Theorem~\ref{thm:divergence simple 8}.

\section{Complete Proofs of Section~\ref{sec:pos and upos}}
\label{sec: proof sec pos and upos}
\subsection{Proof of Theorem~\ref{thm: ultimate pos non hom any order}}
\label{app: proof of ultimate pos non hom any order}
Let \(\mathbf{v}\) be a simple, non-degenerate, inhomogeneous LRS, and
consider the homogeneous LRS \(\mathbf{u}=\homLRS(\mathbf{v})\). If
\(\mathbf{u}\) is a simple LRS, then by~\cite{OW14a} we can effectively
decide its Ultimate Positivity. We assume henceforth that \(\mathbf{u}\)
is not simple.
	
By Property~\ref{prop:homogenization eigenvalues}, it follows that the
characteristic roots of \(\mathbf{u}\) all have multiplicity \(1\), apart
from the characteristic root \(1\) which has multiplicity \(2\). Consider
the dominant modulus \(\rho\) of \(\mathbf{u}\). If \(\rho>1\), then by
writing the exponential polynomial of \(\mathbf{u}\), we have
	\begin{equation}
	\label{eq: exp poly effective ult pos any order}
	\frac{u_n}{\rho^n}=a+\sum_{i=1}^m (c_i\lambda_i^n+\conj{c_i}\conj{\lambda_i^n})+r(n)
	\end{equation}
	with \(a\in \RR\), \(c_i\in \CC\setminus \RR\) and \(|\lambda_i|=1\)
        for every \(1\le i\le m\), and \(|r(n)|\) exponentially decaying.
        Crucially, since \(1\) is not a dominant characteristic root,
        its effect is enveloped in \(r(n)\). Specifically, we observe
        that the analysis of effective Ultimate Positivity
        in~\cite{OW14c} only relies on Proposition~\ref{prop: growth
          r(n)}. Since this holds in the case at hand, we can
        effectively decide Ultimate Positivity when \(1\) is not a
        dominant characteristic root.
	
	Finally, if \(1\) is a dominant characteristic root, the
        exponential polynomial of \(\mathbf{u}\) can be written as
	\begin{equation}
	\label{eq: exp poly effective ult pos 1 dominant any order}
	u_n=A(n)+\sum_{i=1}^m (c_i\lambda_i^n+\conj{c_i}\conj{\lambda_i^n})+r(n).
	\end{equation}
	We observe that in this case, \(u_n\) is ultimately positive iff
        it diverges (indeed, clearly \(|u_n|\to \infty\)). Thus, we can
        reduce the problem to divergence, and proceed with the
        analysis as in Section~\ref{sec: divergence} Case~\ref{case:
          rho=1 A not const every C const}.  \qed

\subsection{Proof of Theorem~\ref{thm: pos non hom 8}}
\label{app: proof of pos non hom 8}
Given the proof of Theorem~\ref{thm: effective ult pos non hom 8},
Positivity is now easily decidable: given an inhomogeneous simple LRS
\(\mathbf{u}\) of order at most \(8\), decide if its ultimately positive,
and if so - compute the bound from which it is ultimately
positive. Then deciding Positivity amounts to checking a finite
number of elements.
	
Note that the bound computed in Theorem~\ref{thm: effective ult pos
  non hom 8} is \(N=2^{\norm{\mathbf{u}}^{O(1)}}\). This implies that checking
whether an ultimately-positive LRS is \emph{not} positive can be done
using a \emph{guess-and-check} procedure, and employing \(\posslp\) in
order to compute double exponential numbers. This yields an
{\NPposslp} algorithm. Thanks to~\cite{Allender}, we obtain an upper
bound of \(\mathbf{coNP}^{\mathbf{PP}^{\mathbf{PP}^{\mathbf{PP}}}}\) for
Positivity (see~\cite{OW14a} for details).  \qed
%%
%% Bibliography
%%

%% Please use bibtex, 

\end{document}